\documentclass[aps,10pt,twocolumn,superscriptaddress]{revtex4-1}

\usepackage{times}
\usepackage{amsmath,amssymb,bm,graphicx,xcolor}
\usepackage[breaklinks,colorlinks=true,urlcolor=blue,citecolor=blue,linkcolor=blue]{hyperref}

\begin{document}
\title{Signatures of gapless fermionic spinons on a strip of the kagome Heisenberg antiferromagnet}

\date{\today}

\author{Amir M-Aghaei}
\affiliation{Department of Physics and Astronomy, University of California at Riverside, Riverside, California 92521, USA}

\author{Bela Bauer}
\affiliation{Station Q, Microsoft Research, Santa Barbara, California 93106, USA}

\author{Kirill Shtengel}
\affiliation{Department of Physics and Astronomy, University of California at Riverside, Riverside, California 92521, USA}

\author{Ryan V. Mishmash}
\affiliation{Department of Physics and Institute for Quantum Information and Matter, California Institute of Technology, Pasadena, California 91125, USA}
\affiliation{Walter Burke Institute for Theoretical Physics, California Institute of Technology, Pasadena, California 91125, USA}
\affiliation{Department of Physics, Princeton University, Princeton, New Jersey 08540, USA}

\begin{abstract}
The search for exotic quantum spin liquid states in simple yet realistic spin models remains a central challenge in the field of frustrated quantum magnetism. Here we consider the canonical nearest-neighbor kagome Heisenberg antiferromagnet restricted to a quasi-1D strip consisting entirely of corner-sharing triangles. Using large-scale density matrix renormalization group calculations, we identify in this model an extended gapless quantum phase characterized by central charge $c=2$ and power-law decaying spin and bond-energy correlations which oscillate at tunably incommensurate wave vectors.  We argue that this intriguing spin liquid phase can be understood as a marginal instability of a two-band spinon Fermi surface coupled to an emergent U(1) gauge field, an interpretation which we substantiate via bosonization analysis and Monte Carlo calculations on model Gutzwiller variational wave functions.  Our results represent one of the first numerical demonstrations of emergent fermionic spinons in a simple SU(2) invariant nearest-neighbor Heisenberg model beyond the strictly 1D (Bethe chain) limit.
\end{abstract}

\maketitle

Beginning with Anderson's seminal proposal of the resonating valence bond state~\cite{Anderson73_MatResBull_8_153, Anderson87_Science_235_1196}, physicists have been actively searching for exotic ground states of spin-1/2 quantum antiferromagnets for more than four decades~\cite{Balents10_Nature_464_199, Savary17_QSLreview_RepProgPhys_80_016502, Kanoda17_QSLreview_RMP_89_025003}. While there have been numerous theoretical and numerical sightings of such quantum spin liquid states over the years, the most convincing demonstrations have typically required going beyond the simplest SU(2) invariant nearest-neighbor Heisenberg model---examples of success include quantum dimer models~\cite{Rokhsar88_PRL_61_2376, Moessner01_PRL_86_1881} or spin models with some combination of, for example, extended two-spin interactions, spin-exchange anisotropy, special conservation laws, and/or multi-site ring-exchange interactions~\cite{Balents02_BFG_PRB_65_224412, Motrunich02_FractionalizedModels_PRB_66_205104, Kitaev06_AnnPhys_321_2, Yao07_ExactCSL_PRL_99_247203, Tay10_PRL_105_187202, Melko11_BHspinLiquid_NatPhys_7_772, Bauer14_ChiralKagome_NatureCommun_5_5137, Sheng14_KagomaCSL_ScientificReports_4_6317, Motrunich05_PRB_72_045105, Block11_PRL_106_157202}.

\begin{figure}[b]
  \centering
  \includegraphics[width=0.88\columnwidth]{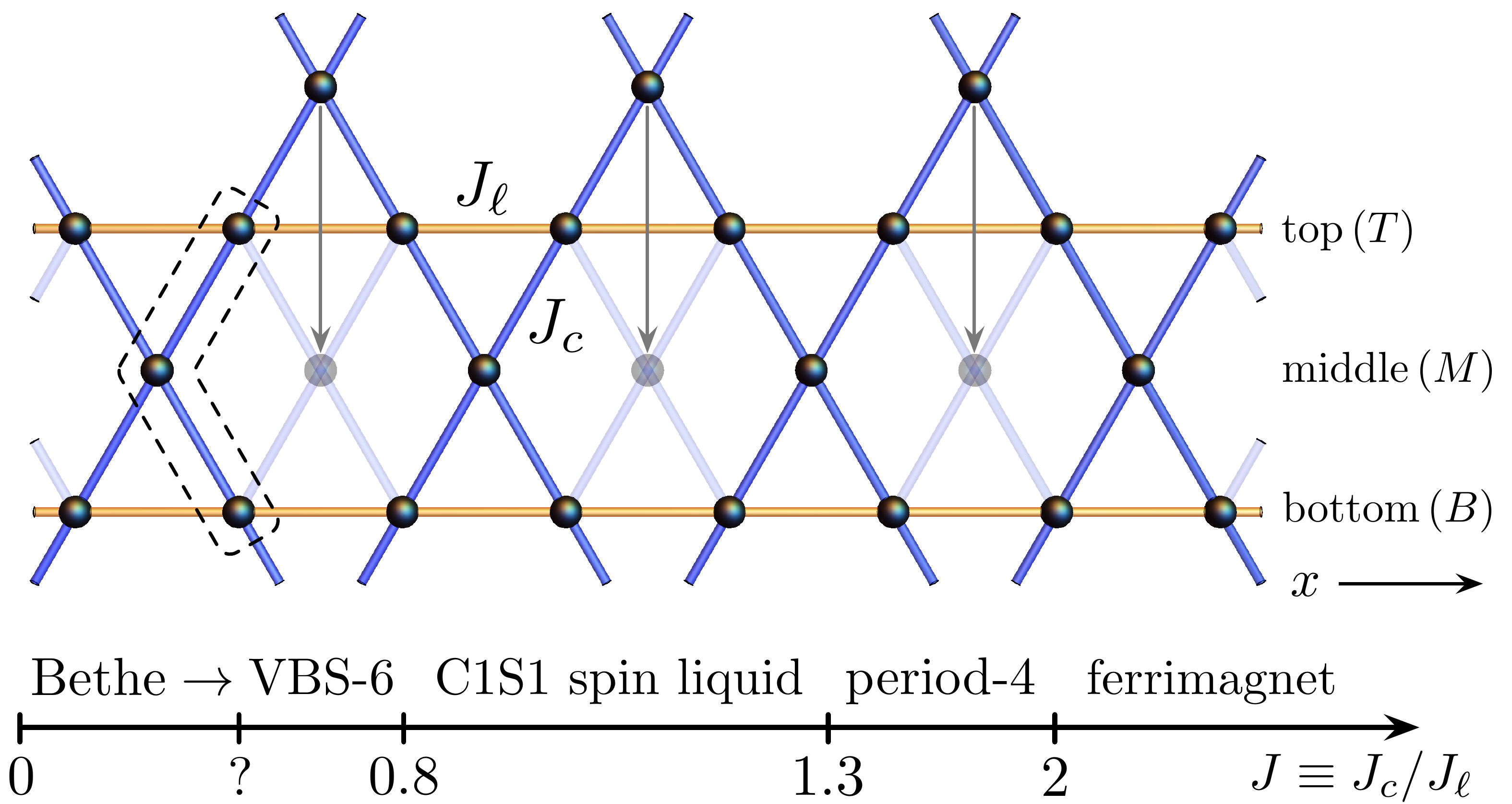}
  \caption{The kagome strip ladder (top) and its numerically obtained phase diagram (bottom). In this work, we imagine ``wrapping'' the lattice such that the topmost row of sites lies on the middle chain (see arrows); the resulting three-site unit cell is boxed by a dashed line. We identify a phase with two 1D gapless modes resulting from gapless bands of fermionic spinons in the regime $0.8\lesssim J\lesssim1.3$.}
  \label{fig:setup}
\end{figure}

One possible exception to this rule is the famous two-dimensional (2D) kagome Heisenberg antiferromagnet, where recent numerical calculations~\cite{White11_Science_332_6034, Depenbrock12_PRL_109_067201} indicate that even the simplest model with SU(2) invariant nearest-neighbor two-spin interactions exhibits spin liquid behavior, a theoretical possibility originally proposed in the early 1990s~\cite{Sachdev92_KagomeSL_PRB_45_12377}. While most of the recent effort (see, for example, Refs.~\cite{White11_Science_332_6034, Depenbrock12_PRL_109_067201, Jiang08_EarlyKagomeDMRG_PRL_101_117203, Jiang12_NaturePhys_8_902, Sheng14_TopoFromDMRG_PRB_89_075110, Kolley15_J1J2Kagome_PRB_91_104418, Sheng15_J1J2J2Kagome_PRB_91_075112, Pollmann17_DiracKagome_PRX_7_031020, Ran16_KagomePEPS_arXiv_1610.02024, Normand17_KagomePESS_PRL_118_137202, Becca11_DiracVMC_PRB_84_020407, Becca13_DiracVMC_PRB_87_060405, Becca14_DiracVMC_PRB_89_020407, Becca15_DiracVMCJ1J2_PRB_91_020402, Becca16_DiracComment_arXiv_1606.02255, changlani_macroscopically_2018}) on kagome systems has been focused on approaching the 2D limit, there remains a particular quasi-one-dimensional (quasi-1D) version that has remarkably evaded both complete numerical characterization and theoretical understanding: the narrowest wrapping of the kagome lattice on a cylinder that consists purely of corner-sharing triangles (see Fig.~\ref{fig:setup}), i.e., the \emph{kagome strip} \footnote{Note that \emph{kagome strip} has referred to a different lattice in the past~\cite{Tsvelik98_KagomeStrip_PRL_81_1694, Waldtmann00_PRB_62_9472}. Our ladder has also been called the \emph{three-spin ladder}~\cite{Waldtmann00_PRB_62_9472} or the \emph{bow-tie ladder}~\cite{Brenig08_BowTieHoles_JPhysCondMat_20_415204}.}. Below, we study the nearest-neighbor spin-1/2 Heisenberg model,
\begin{equation}
H = \sum_{\langle i,j\rangle} J_{ij}\,\mathbf{S}_i \cdot \mathbf{S}_j,
\label{eq:Heis}
\end{equation}
on this lattice with antiferromagetic leg and cross couplings $J_\ell=1$ and $J_c\equiv J\ge 0$, respectively (see Fig.~\ref{fig:setup}). For $J=0$, the model consists of two decoupled Bethe chains (with free spins in the middle chain), while for $J\to\infty$ the model is bipartite and exhibits a conventional ferrimagnetic phase~\cite{Lieb62_LiebMattisFerri_JMathPhys_3_749, Waldtmann00_PRB_62_9472}. Our main interest is in the region $0.8\lesssim J\lesssim2.0$, where in an early study Waldtmann et al.~\cite{Waldtmann00_PRB_62_9472} provided numerical evidence for a gapless ground state but were unable to fully clarify its nature \footnote{Similar conclusions were also reached more recently by L{\"u}scher, McCulloch, and L{\"a}uchli~\cite{Lauchli}.}.

Our main finding is that for $0.8\lesssim J\lesssim1.3$ this model harbors an exotic phase with $c=2$ gapless modes and power-law spin correlations and bond-energy textures which oscillate at incommensurate wave vectors tunable by $J$. We will argue that this phase---which respects all symmetries, including lattice translations and time reversal---can be understood as a marginal instability of a two-band U(1) spinon Fermi surface state, i.e., ``spin Bose metal'' (SBM)~\cite{Sheng09_PRB_79_205112}, on this kagome strip. (Unlike in the U(1) Dirac spin liquid 
\cite{Hastings00_PRB_63_014413, Hermele04_PRB_70_214437, Ran07_PRL_98_117205, Hermele08_PRB_77_224413, Becca11_DiracVMC_PRB_84_020407, Becca13_DiracVMC_PRB_87_060405, Becca14_DiracVMC_PRB_89_020407, Becca15_DiracVMCJ1J2_PRB_91_020402, Becca16_DiracComment_arXiv_1606.02255,Pollmann17_DiracKagome_PRX_7_031020}, 
the spinons in our state see zero flux.) The spinon Fermi surface state has been considered before~\cite{Ran07_PRL_98_117205, Marston08_PRL_101_027204} in the context of the 2D kagome antiferromagnet and its associated prototypical experimental realization herbertsmithite (see Ref.~\cite{Norman16_HerbertsmithiteReview_RMP_88_041002} for a review); however, it is most famous as a proposed theory for several triangular-lattice spin-liquid materials~\cite{Motrunich05_PRB_72_045105, Lee05_PRL_95_036403, Chen16_YbMgGaO4SBM_Nature_540_559}.  It is quite remarkable that a simple model such as this quasi-1D descendant of the nearest-neighbor kagome antiferromagnet gives rise to the exotic physics of multiple bands of fermionic spinons:  While it is well-known that one such band can faithfully describe the Bethe chain phase of the 1D Heisenberg model~\cite{Haldane88_HaldaneShastry_PRL_60_635, Shastry88_HaldaneShastry_PRL_60_639}, other numerically well-established realizations of emergent gapless fermionic slave particles beyond strictly 1D have typically required complicated interactions in the Hamiltonian~\cite{Sheng08_PRB_78_054520, Sheng09_PRB_79_205112, Block11_PRL_106_046402, Block11_PRL_106_157202, Mishmash11_PRB_84_245127, Jiang13_Nature_493_39}.

\begin{figure}[t]
  \centering
  \includegraphics[width=0.8\columnwidth]{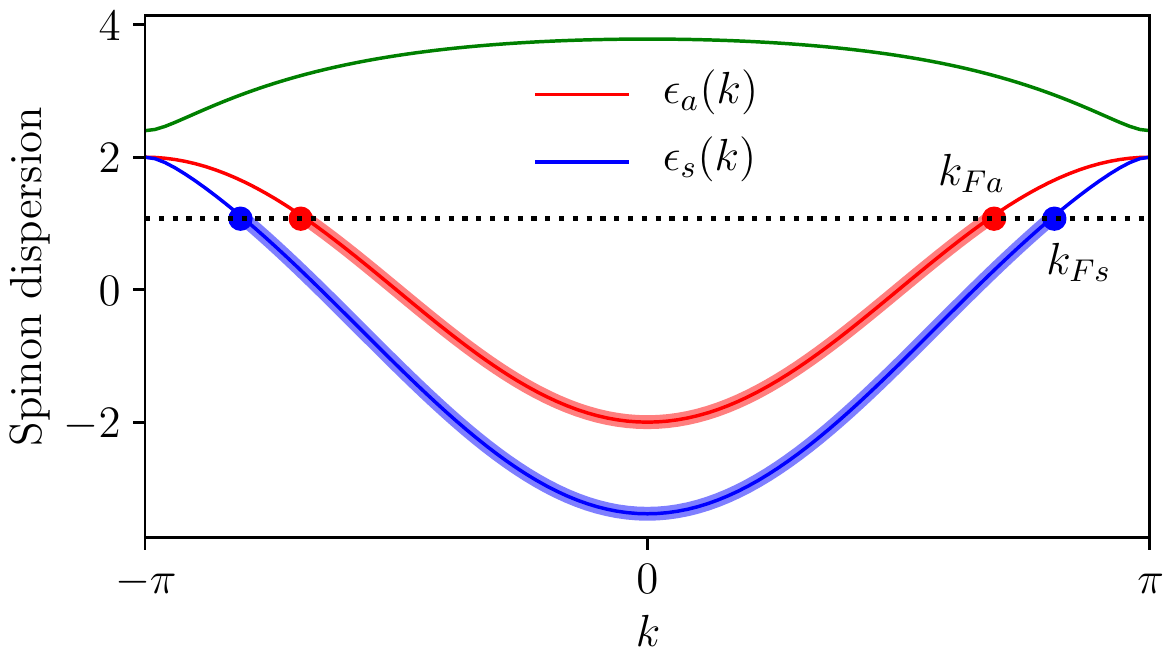}
  \caption{Characteristic spinon band structure for states with $\mu<0$ (here $t_{c}=1.0$, $\mu=-2.4$). There are two partially filled 1D bands, one symmetric ($s$) and one antisymmetric ($a$) under leg interchange. The DMRG ground state for $0.8\lesssim J\lesssim1.3$ on the kagome strip can be well-described as follows: (1) take this $c=4$ mean-field state, (2) include gauge fluctuations, and (3) gap out the spin mode $\theta_{s\sigma}$ for the symmetric band, thereby producing a C1S1 spin liquid state with $c=2$.}
  \label{fig:bands}
\end{figure}

For our theoretical formalism, we take the standard approach~\cite{Wen04_MBQFT} of describing spin liquid states by decomposing the physical spin-1/2 operator $\mathbf{S}_i$ in terms of fermionic spinons $f_{i\alpha}$ subject to the microscopic constraint of one spinon per site, i.e., $\mathbf{S}_i = \frac{1}{2}\sum_{\alpha,\beta=\uparrow,\downarrow}f_{i\alpha}^\dagger\bm{\sigma}_{\alpha\beta}f_{i\beta}$ with $\sum_{\alpha}f_{i\alpha}^\dagger f_{i\alpha} = 1$. We consider a mean-field ansatz for the spinons with nearest-neighbor hopping strengths of $t_\ell = 1$ on the legs and two (real) free parameters, the nearest-neighbor cross-bond hopping $t_{c}$ and the on-site chemical potential $\mu$ on the (vertically) middle sites (see Fig.~\ref{fig:setup}). We only consider unpolarized spin-singlet states so that each spin species is exactly at half filling. A representative spinon band structure for this ansatz is shown in Fig.~\ref{fig:bands}. There are three 1D bands: the topmost and bottommost bands have wave functions symmetric (``$s$'') under interchange of the top and bottom legs, while the middle band's wave functions are antisymmetric (``$a$'') under this symmetry.  We will focus on the case $\mu<0$, which leads to partial filling of the lowest two bands (see Fig.~\ref{fig:bands}), hence producing a state with $c=4$ (two spin \& two charge) gapless modes at the mean-field level.


To go beyond mean field, we couple the spinons to a U(1) gauge field. While the corresponding 2D theory of coupling a Fermi surface to a U(1) gauge field is notoriously challenging~\cite{Lee06_RevModPhys_78_17, Lee08_PRB_78_085129, Lee09_PRB_80_165102, Metlitski10_PRB_82_075127}, including U(1) gauge fluctuations at long wavelengths along a quasi-1D ladder can be readily achieved via bosonization~\cite{Sheng08_PRB_78_054520, Sheng09_PRB_79_205112, Giamarchi03_1D}.
Specifically, integrating out the gauge field produces a mass term for the particular linear combination of bosonized fields corresponding to the overall (gauge) charge mode $\theta_{\rho+}$, thus implementing a coarse-grained version of the on-site constraint mentioned above.
For the two-band situation depicted in Fig.~\ref{fig:bands}, the resulting theory is a highly unconventional $c=3$ Luttinger liquid with one gapless (``relative'') charge mode $\theta_{\rho-}$ and two gapless spin modes $\theta_{s \sigma}$ and $\theta_{a \sigma}$, i.e., a C1S2 SBM state (where C$\alpha$S$\beta$ denotes a state with $\alpha$ ($\beta$) gapless charge (spin) modes~\cite{Balents96_PRB_53_12133, Lin97_PRB_56_6569}). In what follows, we present evidence that the kagome strip Heisenberg model realizes a particular instability of the SBM in which one of the two spin modes is gapped while $c=2$ gapless modes remain: a C1S1 state.


We perform large-scale density matrix renormalization group (DMRG) calculations on Eq.~\eqref{eq:Heis} \footnote{We employ the DMRG-MPS routines from the ALPS package~\cite{ALPS11_JStatMech_05_P05001,Dolfi2014}.} and compare these results to variational Monte Carlo (VMC) calculations~\cite{Gros89_AnnPhys_189_53, Ceperley77_PRB_16_3081} on Gutzwiller-projected wave functions based on the above SBM theory. While our VMC calculations oftentimes provide a semiquantitative description of the DMRG data, we mainly use VMC as a cross-check on the analytic theory and to demonstrate that simple---albeit exotic---wave functions can qualitatively describe the intricate behavior observed in the DMRG.  We work on ladders of length $L$ in the $x$ direction and employ both open and periodic boundary conditions (see Appendix~\ref{sec:details}).

We begin with calculations of bond-energy textures induced by open boundary conditions (OBC)~\cite{Ng96_BetheChainDMRG_PRB_54_9854, Motrunich09_PRB_79_235120}. Specifically, we consider the Fourier transform of local nearest-neighbor spin-spin correlations along the bottom leg: $\mathcal{B}_{q} \equiv \sum_{x}\mathrm{e}^{-iqx} \langle \mathbf{S}^{B}_{x}\cdot\mathbf{S}^{B}_{x+1}\rangle$, where here and in what follows $\mathbf{S}_x^\lambda$ is the spin operator at horizontal position $x$ and vertical position $\lambda=T,M,B$ (for ``top'', ``middle'', and ``bottom''; see Fig.~\ref{fig:setup}).  Such quantities contain content similar to the dimer structure factor~\cite{Motrunich09_PRB_79_235120, Sheng09_PRB_79_205112}, yet are less formidable to compute on large systems.
In Fig.~\ref{fig:bondtextures}, we show DMRG measurements of $\mathcal{B}_{q}$ on an OBC system of length $L=60$
(see Appendix~\ref{sec:details}). We see that $\mathcal{B}_q$ generically shows two prominent features centered symmetrically about wave vector $\pi/2$. These features are power-law singularities for $0.8\lesssim J\lesssim1.3$; we will later discuss the Bragg peaks observed at $J=0.78$.  Defining $q_<$ ($q_>$) as the smaller (larger) wave vector, notice that $q_<$ ($q_>$) increases (decreases) with increasing $J$, but the two wave vectors always satisfy $q_< + q_> = \pi$.

\begin{figure}[t]
  \centering
  \includegraphics[width=0.95\columnwidth]{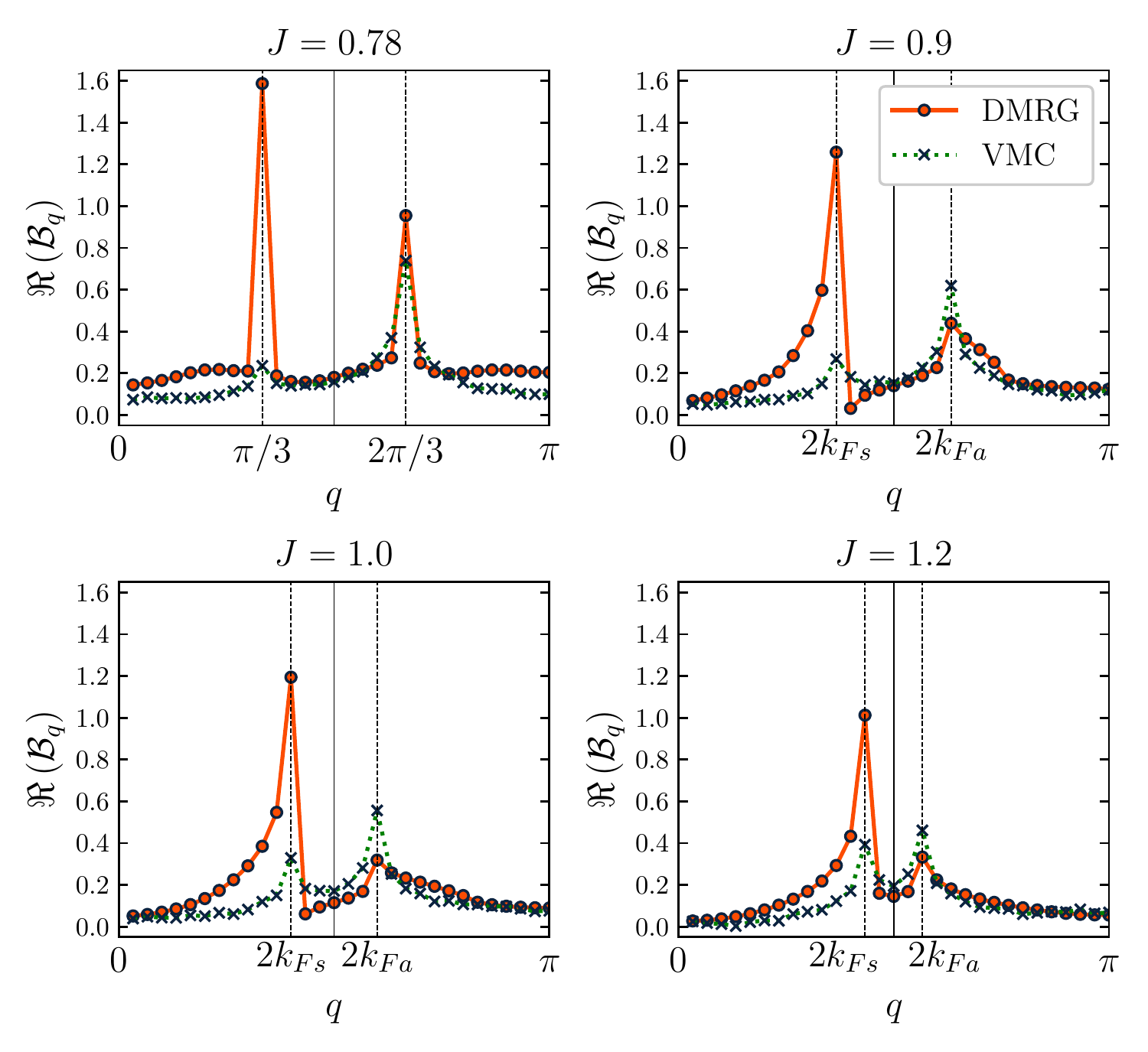}
  \caption{Fourier transform of (leg-bond) bond-energy textures induced by OBC on a length $L=60$ kagome strip at $J = 0.78, 0.9, 1.0$, and $1.2$. We show both DMRG data and VMC data for bare Gutzwiller SBM states. A wave function for the proposed C1S1 state would appear similar to the SBM except it would have a more prominent feature at $q_< = 2k_{Fs}$ due to lowering of the scaling dimension of the associated operator upon pinning of $\theta_{s\sigma}$. At $J=0.78$, the DMRG ground state is a fully gapped period-6 VBS phase (see text). For analogous data of bond-energy textures involving the \emph{cross bonds}, please see Fig.~\ref{fig:crossbonds} in Appendix~\ref{sec:observables}.}
  \label{fig:bondtextures}
\end{figure}

The presence of such power-law singularities at wave vectors tunable by a coupling parameter, yet obeying particular sum rules, is suggestive of multiple bands of gapless fermionic spinons~\cite{Sheng08_PRB_78_054520, Sheng09_PRB_79_205112, Block11_PRL_106_046402, Block11_PRL_106_157202, Mishmash11_PRB_84_245127, Jiang13_Nature_493_39}.  In Fig.~\ref{fig:bondtextures}, we also include VMC calculations on wave functions obtained by Gutzwiller projecting the free fermion states of the form shown in Fig.~\ref{fig:bands}---these are model wave functions for the SBM~\cite{Motrunich05_PRB_72_045105, Sheng09_PRB_79_205112, Block11_PRL_106_157202} (see also Appendix~\ref{sec:vmc}). Such wave functions exhibit power-law singularities in physical quantities at various ``$2k_{F}$'' wave vectors, i.e., wave vectors obtained by connecting sets of Fermi points in Fig.~\ref{fig:bands}. Specifically, for the SBM states considered, we expect and observe features in $\mathcal{B}_q$ at wave vectors $q=2k_{Fs}$ and $2k_{Fa}$, where $2k_{Fs}+2k_{Fa}=\pi\mod2\pi$ due to the half-filling condition. The overall qualitative agreement between VMC and DMRG measurements of $\mathcal{B}_q$ in Fig.~\ref{fig:bondtextures} is notable; recall that the VMC states have only two free parameters. We can now make the following identification with the wave vectors $q_<$ and $q_>$ discussed earlier: $q_< = 2k_{Fs}$ and $q_> = 2k_{Fa}$.


\begin{figure}[t]
  \centering
  \includegraphics[width=0.95\columnwidth]{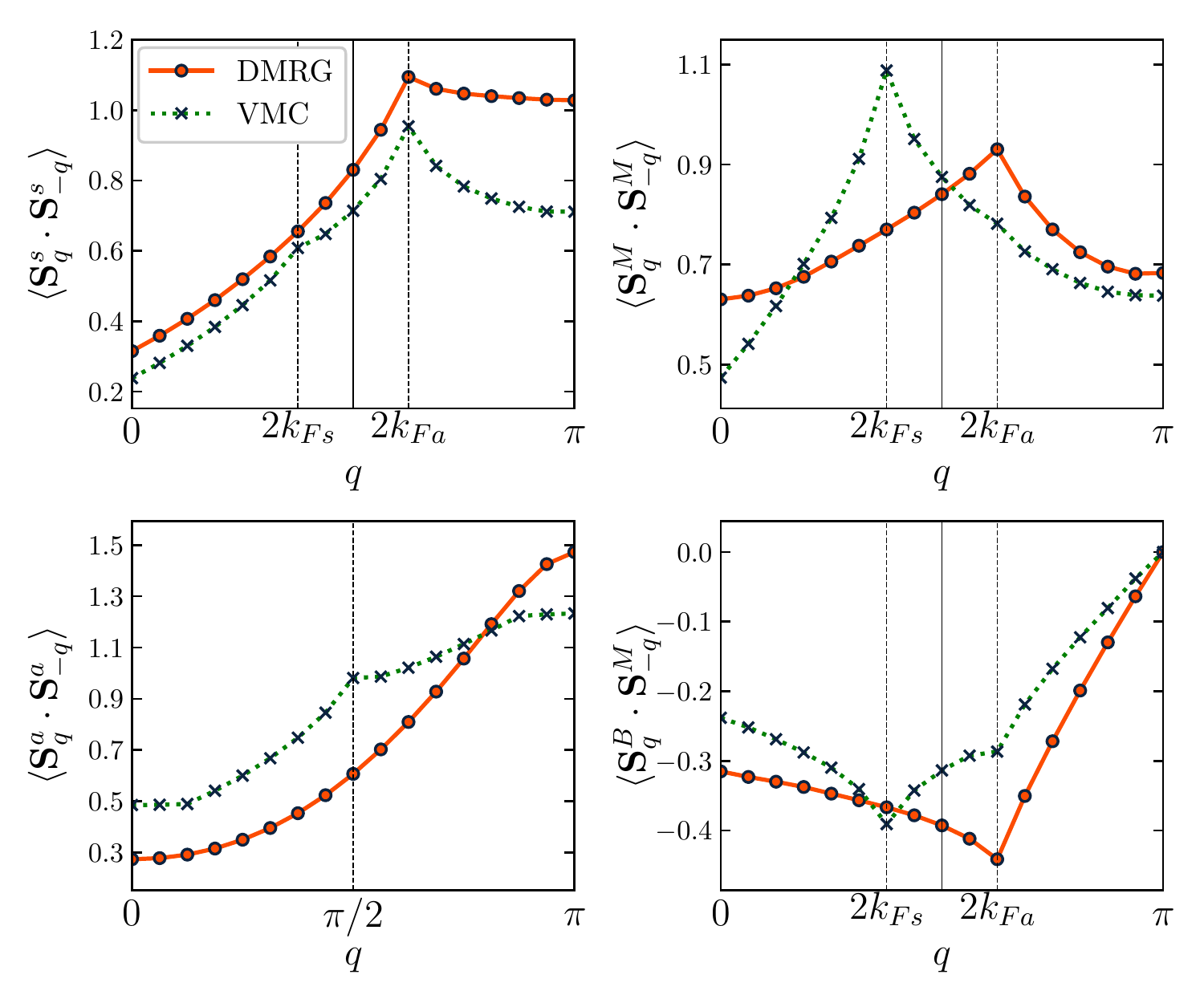}
  \caption{Spin structure factors at $J=0.9$ on a $L=32$ system with PBC (see text for definitions of operators). As in Fig.~\ref{fig:bondtextures}, we show both DMRG and bare Gutzwiller (SBM) VMC calculations. All features at wave vectors $q_< = 2k_{Fs}$ and $\pi/2$ would be absent in a wave function for the proposed C1S1 state---indeed these features are absent in the DMRG data.}
  \label{fig:ssf}
\end{figure}

Next we turn to measurements of the spin structure factor. Defining $\mathbf{S}_x^{s/a}\equiv\frac{1}{\sqrt{2}}\left(\mathbf{S}_x^T \pm \mathbf{S}_x^B\right)$, we consider 1D structure factors obtained by Fourier transforming real-space spin-spin correlation functions composed from the spin operators $\mathbf{S}_x^{s}$, $\mathbf{S}_x^{M}$, and $\mathbf{S}_x^{a}$, i.e., $\langle\mathbf{S}_{q}^{s}\cdot\mathbf{S}_{-q}^{s}\rangle$, $\langle\mathbf{S}_{q}^{M}\cdot\mathbf{S}_{-q}^{M}\rangle$, and $\langle\mathbf{S}_{q}^{a}\cdot\mathbf{S}_{-q}^{a}\rangle$. The former two spin operators are symmetric under leg interchange ($T\leftrightarrow B$), while $\mathbf{S}_x^{a}$ is antisymmetric. To characterize correlations between the outer chains and the middle sites, we also consider the analogous 1D structure factor $\langle\mathbf{S}_{q}^{B}\cdot\mathbf{S}_{-q}^{M}\rangle = \langle\mathbf{S}_{q}^{T}\cdot\mathbf{S}_{-q}^{M}\rangle$. In Fig.~\ref{fig:ssf}, we show DMRG calculations of these four quantities on a system of length $L=32$ with periodic boundary conditions (PBC) at coupling $J=0.9$, which is characteristic of the observed behavior throughout $0.8\lesssim J\lesssim1.3$. As calculated by DMRG, the three structure factors $\langle\mathbf{S}_{q}^{s}\cdot\mathbf{S}_{-q}^{s}\rangle$, $\langle\mathbf{S}_{q}^{M}\cdot\mathbf{S}_{-q}^{M}\rangle$, and $\langle\mathbf{S}_{q}^{B}\cdot\mathbf{S}_{-q}^{M}\rangle$ all reveal clear power-law singularities at a particular incommensurate wave vector $q_> = 10\cdot\frac{2\pi}{32}$, while $\langle\mathbf{S}_{q}^{a}\cdot\mathbf{S}_{-q}^{a}\rangle$ is completely smooth hence indicating exponential decay in real space. Also shown in Fig.~\ref{fig:ssf} are VMC calculations for an appropriate SBM state satisfying $2k_{Fa} = q_>$. As expected, the VMC data shows singular features in $\langle\mathbf{S}_{q}^{s}\cdot\mathbf{S}_{-q}^{s}\rangle$, $\langle\mathbf{S}_{q}^{M}\cdot\mathbf{S}_{-q}^{M}\rangle$, and $\langle\mathbf{S}_{q}^{B}\cdot\mathbf{S}_{-q}^{M}\rangle$ at wave vectors $q_<=2k_{Fs}$ and $q_>=2k_{Fa}$ and in $\langle\mathbf{S}_{q}^{a}\cdot\mathbf{S}_{-q}^{a}\rangle$ at wave vector $\pi/2$. In this case, the qualitative agreement between VMC and DMRG remains intact only near the wave vector $q_> = 2k_{Fa}$: the DMRG data is completely lacking any structure at both $q_< = 2k_{Fs}$ (symmetric cases) and at $\pi/2$ (antisymmetric case).

We can explain in a universal way this discrepancy by postulating that the spin mode $\theta_{s\sigma}$ is gapped in the DMRG state. Indeed, in the low-energy SBM theory, there is an allowed four-fermion single-band $2k_{F}$ backscattering interaction which, upon bosonization, contains a nonlinear cosine potential~\cite{Balents96_PRB_53_12133, Sheng09_PRB_79_205112, Motrunich10_PRB_81_045105, Ko13_KondoC1S1_PRB_87_205107}:
\begin{equation}
V_{ss}^\perp = \lambda_{ss}^\sigma\cos(2\sqrt{2}\theta_{s\sigma}).
\end{equation}
If $\lambda_{ss}^\sigma < 0$, this term is marginally relevant, and the field $\theta_{s\sigma}$ becomes pinned~\cite{Balents96_PRB_53_12133, Sheng09_PRB_79_205112}. Assuming all other allowed interactions are irrelevant or marginally irrelevant, the resulting state is an unconventional C1S1 Luttinger liquid with two gapless modes, $\theta_{\rho-}$ and $\theta_{a\sigma}$, and one nontrivial Luttinger parameter $g_{\rho-} < 2/3$ (see Appendix~\ref{sec:theory} and Ref.~\cite{Sheng09_PRB_79_205112}). Unfortunately, faithfully describing our proposed C1S1 state via projected variational wave functions cannot be done in a straightforward way (see Appendix~\ref{sec:vmc}). However, based on our theoretical understanding, we can be certain that a C1S1 state would resolve all qualitative differences between the (C1S2 SBM) VMC data and the DMRG data in Figs.~\ref{fig:bondtextures} and \ref{fig:ssf}. Firstly, this state would have short-ranged correlations in the spin structure factor measurements at wave vectors $q_<= 2k_{Fs}$ and $\pi/2$, while retaining power-law behavior at $q_> = 2k_{Fa}$---completely consistent with the DMRG data in Fig.~\ref{fig:ssf}. Secondly, since the long-wavelength component of the bond energy at wave vector $2k_{Fs}$ is proportional to $e^{-i\theta_{\rho-}}\cos(\sqrt{2}\theta_{s\sigma})$, the corresponding feature at $q_< = 2k_{Fs}$ in $\mathcal{B}_{q}$ would actually be \emph{enhanced} relative to the SBM upon pinning of $\theta_{s\sigma}$. This indeed occurs in the DMRG data of Fig.~\ref{fig:bondtextures}, where the feature at $q_< = 2k_{Fs}$ in $\mathcal{B}_{q}$ is significantly more pronounced than that at $q_> = 2k_{Fa}$. Finally, as we show in Appendix~\ref{sec:observables}, the \emph{spin chirality structure factor} as obtained by DMRG is featureless at finite wave vectors. While the C1S2 state would exhibit power-law decaying chirality correlations at various finite wave vectors due to \emph{interband} $2k_F$ processes~\cite{Sheng09_PRB_79_205112}, decay at these wavevectors become short-ranged in the C1S1 state with its gapped spin mode $\theta_{s\sigma}$---this is fully consistent with our DMRG findings in Appendix~\ref{sec:observables}. Furthermore, we observe no Bragg peaks in the chirality structure factor measurements thereby allowing us to clearly rule out spontaneous breaking of time-reversal symmetry in this model \footnote{The realized state is thus unrelated to the gapless chiral U(1) spin liquid states studied in Ref.~\cite{bieri_gapless_2015}.}.

We next describe \emph{instabilities} out of the putative C1S1 phase realized in the DMRG for $0.8\lesssim J\lesssim1.3$.  On one side, in a narrow window $0.75\lesssim J\lesssim0.8$, we find a state with (dominant) period-6 long-range valence bond solid (VBS) order---see the Bragg peaks in the DMRG measurements of $\mathcal{B}_q$ at $J=0.78$ in Fig.~\ref{fig:bondtextures}.  Remarkably, this VBS-6 phase can be naturally understood by analyzing the C1S1 theory at the special commensurate point corresponding to $2k_{Fs} = \pi/3$ and $2k_{Fa} = 2\pi/3$.  Here, there exists an additional symmetry-allowed six-fermion umklapp-type interaction which is necessarily relevant with respect to the C1S1 fixed point, thereby providing a natural explanation for the observed VBS state bordering the C1S1.  On the other side, we observe a strong first-order phase transition (and possibly intervening phase) in the region $J\simeq1.3-1.4$ before entering a phase at still larger $J$ with period-4 bond-energy textures (likely) decaying as a power-law.

\begin{figure}[t]
  \centering
  \includegraphics[width=0.9\columnwidth]{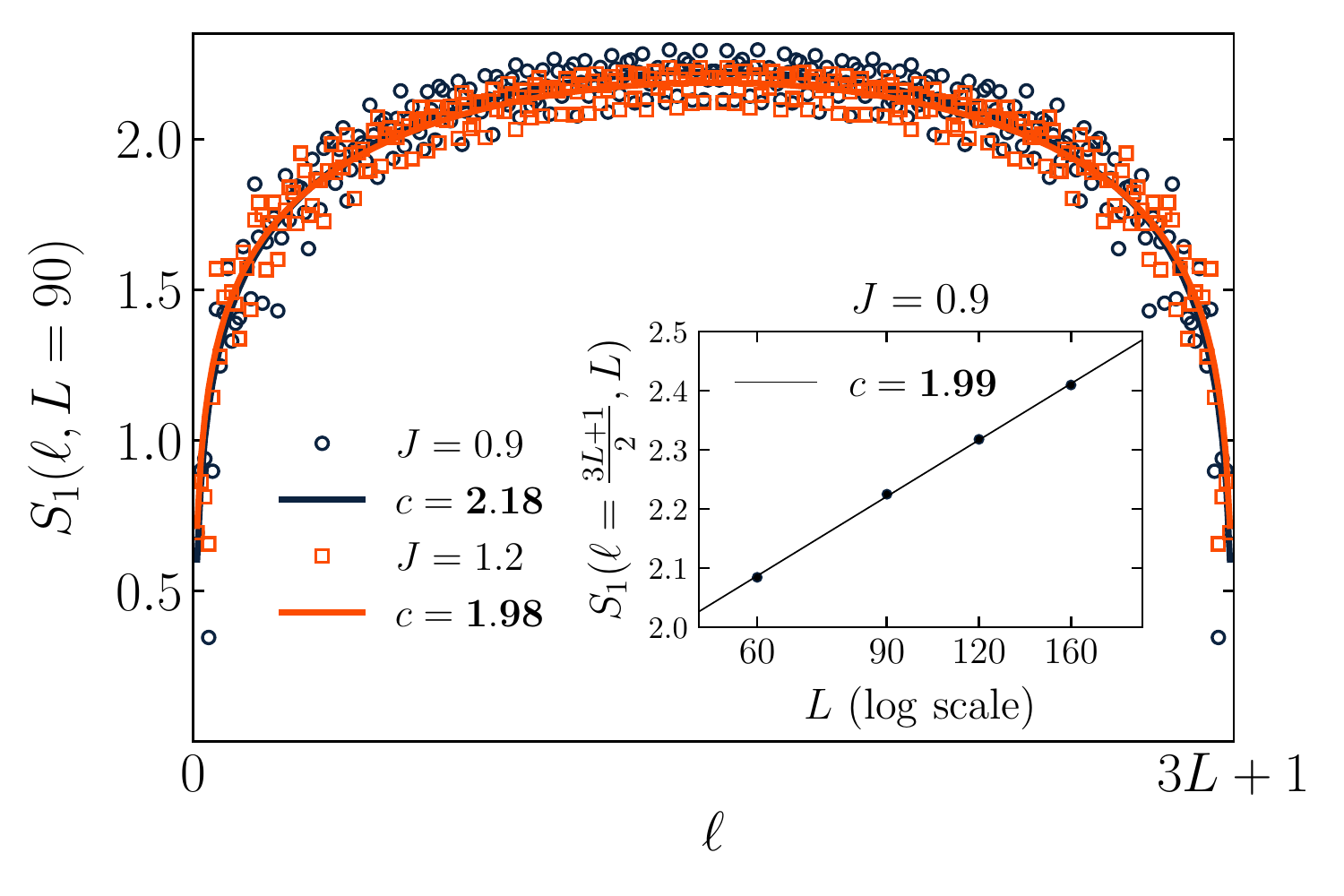}
  \caption{Scaling of the von Neumann entanglement entropy $S_1$ versus subsystem size $\ell$ as calculated by DMRG on an OBC system of length $L=90$ at $J=0.9$ and 1.2. In the inset, we show for $J=0.9$ the mid-system entanglement entropy as we vary $L$. The solid curves are fits to the scaling form~\cite{Cardy04_JStatMech_P06002}, strongly indicating $c=2$ as expected for C1S1.}
  \label{fig:EE}
\end{figure}

We conclude with measurements of the bipartite entanglement entropy, the scaling of which gives access to perhaps the most important universal number characterizing 1D and quasi-1D systems: the central charge $c$, which in our case is equivalent to the number of 1D gapless modes of the realized Luttinger liquid~\cite{Cardy04_JStatMech_P06002}. We perform DMRG calculations on large $x\leftrightarrow-x$ reflection-symmetric OBC systems (see Appendix~\ref{sec:details}) up to length $L=160$ ($3L+1=481$ total sites), and as is clearly evident in Fig.~\ref{fig:EE}, fits to the usual scaling form~\cite{Cardy04_JStatMech_P06002} strongly suggest $c=2$ for $0.8\lesssim J\lesssim1.3$. This is precisely the number of 1D gapless modes expected for the C1S1 state.

In conclusion, we have presented convincing numerical evidence that the ground state of the simple kagome strip Heisenberg model can be described as an intriguing C1S1 spin liquid phase, a marginal instability of the spin Bose metal (i.e., U(1) spinon Fermi surface with no flux) on this ladder. We emphasize that by employing fully controlled numerical and analytical techniques we can understand the realized exotic phase very thoroughly in terms of gapless fermionic spinons---indeed the ability to develop such a complete understanding of an exotic phase of matter in a simple nearest-neighbor Heisenberg spin model is exceedingly rare \footnote{In Appendix~\ref{sec:azaria}, we contrast our results with the $c=2$ fixed point realized in a frustrated three-leg spin ladder in Ref.~\cite{azaria_chiral_1998}.}.
While the simplest Dirac-spin-liquid-like mean-field starting point on this kagome strip (with $\pi$ flux through the hexagons in Fig.~\ref{fig:setup}) leads to a fully gapped state at the mean-field level, it would be interesting to search for other possible two-band scenarios with the hope of connecting our results to recent work suggesting a gapless state in the 2D kagome Heisenberg antiferromagnet~\cite{Pollmann17_DiracKagome_PRX_7_031020, Ran16_KagomePEPS_arXiv_1610.02024, Normand17_KagomePESS_PRL_118_137202}.  More generally, it is interesting to ask \emph{why} a state such as the C1S1 would be realized in our model: Previous realizations of the spin Bose metal
itself involved interactions appropriate for \emph{weak Mott insulators} with substantial charge fluctuations~\cite{Sheng09_PRB_79_205112, Block11_PRL_106_157202, Mishmash15_PRB_91_235140}, while the simple Heisenberg model of our work is appropriate only in the strong Mott regime. Perhaps our work can thus give some guidance on realizing exotic spin liquid states with emergent fermionic spinons in simple models of frustrated quantum antiferromagnets.

\begin{acknowledgments}
We are very grateful to Andreas L{\"a}uchli for discussions and for sharing with us his unpublished DMRG results on the same model~\cite{Lauchli} and also to Olexei Motrunich for many useful discussions on this work. This work was supported by the NSF through Grant No.~DMR-1411359 (A.M.A. and K.S.); the Walter Burke Institute for Theoretical Physics at Caltech; and the Caltech Institute for Quantum Information and Matter, an NSF Physics Frontiers Center with the support of the Gordon and Betty Moore Foundation.
\end{acknowledgments}

\appendix

\section{Details of the DMRG calculations and additional data} \label{sec:details}

\begin{figure}[t]
  \centering
  \includegraphics[width=0.9\columnwidth]{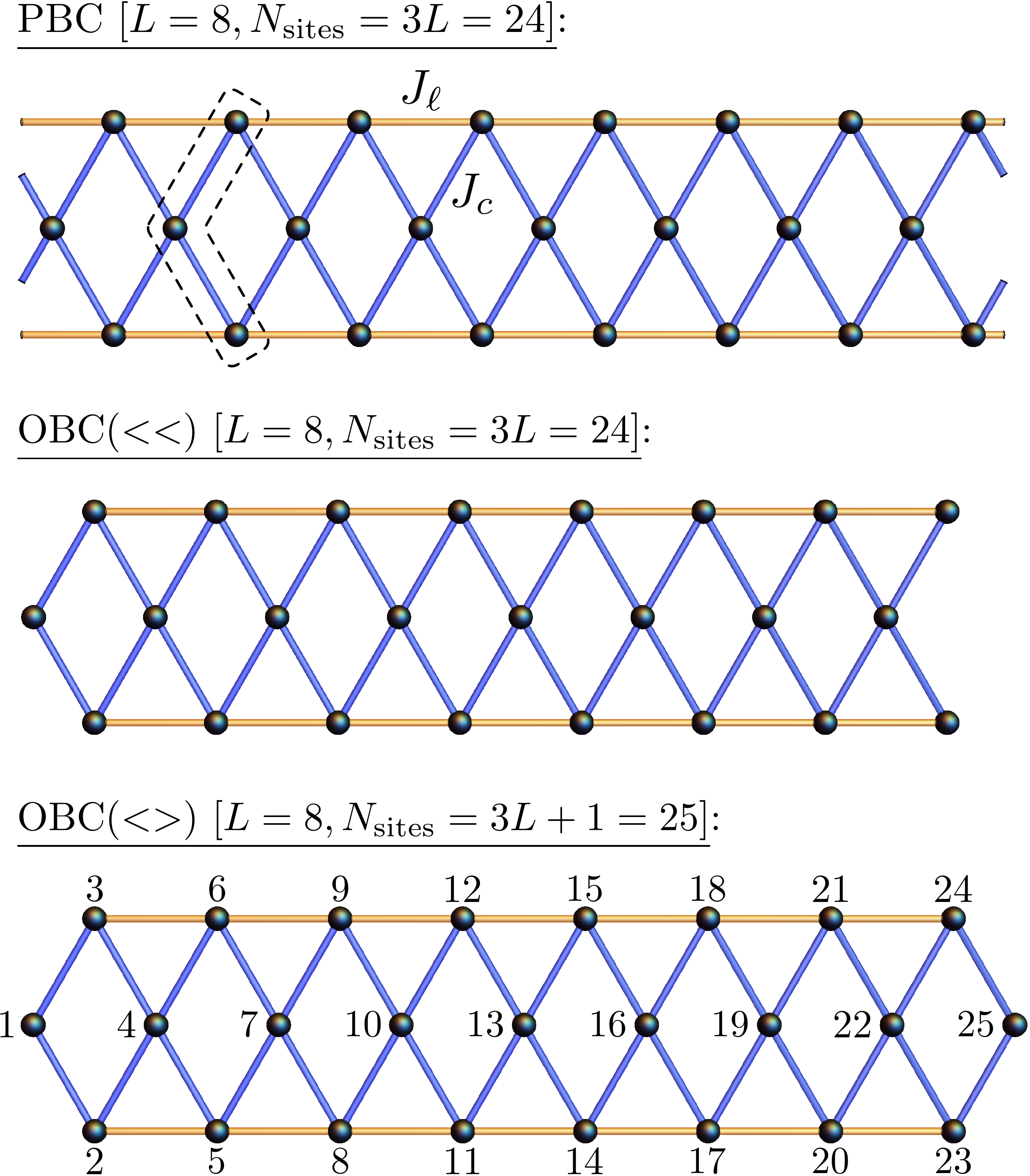}
  \caption{Kagome strip clusters with different boundary conditions (from top to bottom): PBC, OBC($<<$), and OBC($<>$). The relation to the 2D kagome lattice structure is illustrated in Fig.~\ref{fig:setup}. The ``leg'' bonds are orange with associated coupling strength $J_\ell=1$ in the Hamiltonian, while the ``cross'' bonds are blue with coupling $J_c\equiv J\geq0$. In each case, the example lattice corresponds to a length $L=8$ system. The site labels in the OBC($<>$) case indicate the progression of subsystem bipartitions used in our entanglement entropy analysis (see Fig.~\ref{fig:EE}).}
  \label{fig:clusters}
\end{figure}

We perform large-scale DMRG calculations on the kagome strip Heisenberg model [see Eq.~\eqref{eq:Heis}] for finite-size systems with either periodic (PBC) or open (OBC) boundary conditions in the $x$ direction. The precise lattice geometries we use are shown in Fig.~\ref{fig:clusters}. For the PBC setup, a unit cell (of which there are $L$) is boxed by a dashed line. For OBC systems, we consider two different setups, OBC($<<$) and OBC($<>$), where the direction of the two angle brackets indicates the type of boundary termination at the left and right ends of the ladder (see Fig.~\ref{fig:clusters}). Note that the OBC($<>$) configuration exhibits $x\leftrightarrow-x$ reflection symmetry about the centermost site, while OBC($<<$) does not. In all cases, $L$ refers to the number of sites along the bottom (top) chain so that the total number of sites is $N_\mathrm{sites}=3L$ for both PBC and OBC($<<$), while $N_\mathrm{sites}=3L+1$ for OBC($<>$).

For our DMRG simulations, we generally retain a bond dimension of between about $m=$ 1,600 and 4,000 states and perform about 10 to 30 finite-size sweeps, resulting in a density matrix truncation error of $10^{-6}$ or smaller.  All measurements are converged to an accuracy of the order of the symbol size or smaller in the presented plots.

In the main text, we focused on measurements of (1) bond-energy textures, (2) spin structure factors, and (3) bipartite entanglement entropy. Throughout, we define $\mathbf{S}_x^{\lambda}$ as the spin operator at horizontal position $x$ and vertical position $\lambda=T,M,B$ (for the ``top'', ``middle'', and ``bottom'' rows of sites; see Fig.~\ref{fig:setup}). We also define symmetric and antisymmetric combinations of $\mathbf{S}_x^{T}$ and $\mathbf{S}_x^{B}$:
\begin{equation}
\mathbf{S}_x^{s/a}\equiv\frac{1}{\sqrt{2}}\left(\mathbf{S}_x^T \pm \mathbf{S}_x^B\right).
\end{equation}

For the bond-energy texture calculations, we employ OBC and compute the Fourier transform of the nearest-neighbor bond-energy expectation value along one of the horizontal legs (say the bottom chain):
\begin{equation}
\mathcal{B}_{q} \equiv \mathcal{B}_{q}^\mathrm{leg} \equiv \sum_{x=1}^L\mathrm{e}^{-iqx} \langle \mathbf{S}^{B}_{x}\cdot\mathbf{S}^{B}_{x+1}\rangle.
\label{eq:Bqdef}
\end{equation}
For both OBC configurations, a system of length $L$ has $L$ sites---and thus $L-1$ bonds---along the bottom chain. Thus, we define $\langle\mathbf{S}^{B}_{L}\cdot\mathbf{S}^{B}_{L+1}\rangle\equiv0$ when computing $\mathcal{B}_q^\mathrm{leg}$ in Eq.~\eqref{eq:Bqdef} so that $\langle\mathbf{S}^{B}_{x}\cdot\mathbf{S}^{B}_{x+1}\rangle$ is effectively $L$-periodic [for OBC($<<$) in practice we append the 0 to the \emph{beginning} of the real-space vector, $\langle\mathbf{S}^{B}_{0}\cdot\mathbf{S}^{B}_{1}\rangle\equiv0$, before performing the Fourier transform]. Below in Figs.~\ref{fig:crossbonds} and \ref{fig:VBSs}, we present additional data on the analogous (parallel) cross-bond bond-energy textures:
\begin{equation}
\mathcal{B}_{q}^\mathrm{cross} \equiv \sum_{x=1}^L\mathrm{e}^{-iqx} \langle \mathbf{S}^{B}_{x}\cdot\mathbf{S}^{M}_{x-\frac{1}{2}}\rangle.
\label{eq:Bqcrossdef}
\end{equation}
Since the real-space data used to generate $\mathcal{B}_{q}^\mathrm{leg/cross}$ does not generally exhibit $x\leftrightarrow-x$ symmetry [e.g., due to use of OBC($<<$)], our Fourier-space data is in general complex.  For simplicity, we thus plot only the real part: $\Re\,(\mathcal{B}_{q}^\mathrm{leg/cross})$.  Finally, we have confirmed that using OBC($<<$) versus OBC($<>$) does not make a qualitative difference in these bond-energy texture calculations; for presentation in Fig.~\ref{fig:bondtextures} and in Fig.~\ref{fig:crossbonds} below, we use the OBC($<<$) setup.

For the spin structure factor calculations, we use PBC and compute the following four momentum-space spin-spin correlation functions:
\begin{align}
\langle\mathbf{S}_{q}^{s}\cdot\mathbf{S}_{-q}^{s}\rangle & \equiv \frac{1}{L}\sum_{x,x'} e^{-iq(x-x')} \langle\mathbf{S}_x^{s}\cdot\mathbf{S}_{x'}^{s}\rangle, \label{eq:SsSs} \\
\langle\mathbf{S}_{q}^{M}\cdot\mathbf{S}_{-q}^{M}\rangle & \equiv \frac{1}{L}\sum_{x,x'} e^{-iq(x-x')} \langle\mathbf{S}_x^{M}\cdot\mathbf{S}_{x'}^{M}\rangle, \label{eq:SMSM} \\
\langle\mathbf{S}_{q}^{a}\cdot\mathbf{S}_{-q}^{a}\rangle & \equiv \frac{1}{L}\sum_{x,x'} e^{-iq(x-x')} \langle\mathbf{S}_x^{a}\cdot\mathbf{S}_{x'}^{a}\rangle, \label{eq:SaSa} \\
\langle\mathbf{S}_{q}^{B}\cdot\mathbf{S}_{-q}^{M}\rangle & \equiv \frac{1}{L}\sum_{x,x'} e^{-iq(x-x')} \langle\mathbf{S}_x^{B}\cdot\mathbf{S}_{x'}^{M}\rangle. \label{eq:SBSM}
\end{align}

When using PBC, we must necessarily work on smaller systems due to its well-known convergence problems in the DMRG (the largest PBC system presented in this work is for $L=32$, i.e., $N_\mathrm{sites}=96$ total spins).  Within the putative C1S1 state, for $1.0\lesssim J\lesssim1.3$ a relatively small bond dimensions of $m=$ 3,000 results in a converged and almost translationally invariant system, while for $0.8\lesssim J\lesssim1.0$ a perfectly translationally invariant ground state is difficult to achieve even for $m$ as large as 4,800.  In principle, this can be an artifact of finite-momentum in the ground-state wave function~\cite{Sheng09_PRB_79_205112}.  Another culprit could be the near-ordering tendencies of the state at wave vector $q_<$ in the bond energy (see Fig.~\ref{fig:bondtextures}).

At the specific point $J=0.9$, on smaller PBC systems of length $L=18,20,24$, we were able to eventually converge to a translationally invariant state by increasing $m$ and the number of sweeps.  In all of these cases, when measured for a stable but not fully translationally invariant system, we can confirm that measurement of the spin structure factors in Eqs.~\eqref{eq:SsSs}--\eqref{eq:SBSM} (which effectively average $L$ one-dimensional Fourier transforms over all ``origins'' of the system) are identical to those performed on the final translationally invariant states.  Hence, we are confident that the final spin structure factor measurements such as those presented in Fig.~\ref{fig:ssf} are fully converged, accurate representations of the spin correlations in the ground-state wave function.

Below in Appendix~\ref{sec:observables}, we present additional data on spin chirality structure factor measurements, also obtained with PBC. Specifically, we calculate
\begin{align}
\langle\chi_q^B \chi_{-q}^B\rangle & \equiv \frac{1}{L}\sum_{x,x'} e^{-iq(x-x')}\langle\chi_x^B\chi_{x'}^B\rangle, \label{eq:chichiBB} \\
\langle\chi_q^B \chi_{-q}^T\rangle & \equiv \frac{1}{L}\sum_{x,x'} e^{-iq(x-x')}\langle\chi_x^B\chi_{x'}^T\rangle, \label{eq:chichiBT}
\end{align}
where
\begin{equation}
\chi_x^{B/T} \equiv \mathbf{S}_x^{B/T} \cdot (\mathbf{S}_{x+\frac{1}{2}}^M \times \mathbf{S}_{x+1}^{B/T}). \label{eq:chix}
\end{equation}
For simplicity, we take the convention that the real-space two-point correlation functions $\langle\chi_x^B\chi_{x'}^B\rangle$, $\langle\chi_x^B\chi_{x'}^T\rangle$ are zero if the two chirality operators share any common sites.

For our entanglement entropy calculations, we present data on the $x\leftrightarrow-x$ reflection-symmetric OBC($<>$) system. We use a progression of bipartitions as indicated by the site labels in the bottommost panel of Fig.~\ref{fig:clusters}. That is, the first subsystem considered contains the site labeled 1, the second subsystem contains sites 1 and 2, and so on. We compute with DMRG the von Neumann entanglement entropy,
\begin{equation}
S_1(\rho_A) = -\mathrm{Tr}\left(\rho_A\log\rho_A\right),
\end{equation}
where $\rho_A$ is the reduced density matrix for a subsystem $A$. Note that the chosen progression of bipartitions produces data of $S_1$ versus subsystem size $\ell=1,2,\dots,N_\mathrm{sites}-1$ which is symmetric about the middle of the ladder in the $x$ direction. We then perform fits to the calculated entanglement entropy data using the well-known Calabrese-Cardy formula~\cite{Cardy04_JStatMech_P06002} to determine the central charge, $c$. Specifically, we fit to the scaling form
\begin{equation}
S_1(\ell, L) = \frac{c}{6}\log\left(\frac{3L+1}{\pi}\sin\frac{\pi\ell}{3L+1}\right) + A,
\end{equation}
where $3L+1=N_\mathrm{sites}$ is the total number of sites for OBC($<>$). In our fits, we omit $O(10)$ of the smallest/largest subsystems near the ends of the ladder.  The mid-system entanglement entropy data shown in the inset of Fig.~\ref{fig:EE} is simply the raw $S_1$ data for subregions spanning half the system according to the above labeling.  For OBC($<>$) systems with $L$ even ($N_\mathrm{sites}$ odd), as presented in Fig.~\ref{fig:EE}, we must work in the sector with $S^z_\mathrm{tot}=\frac{1}{2}$; we have confirmed that this detail makes no difference in the central charge determination. In addition, we have performed analogous calculations for both PBC and OBC($<<$) systems where pure ``unit-cell bipartitions'' are natural, and we have indeed been able to confirm in those setups as well the result $c=2$ in the putative C1S1 state for $0.8\lesssim J\lesssim1.3$ (data not shown).

\begin{figure}[t]
  \centering
  \includegraphics[width=0.88\columnwidth]{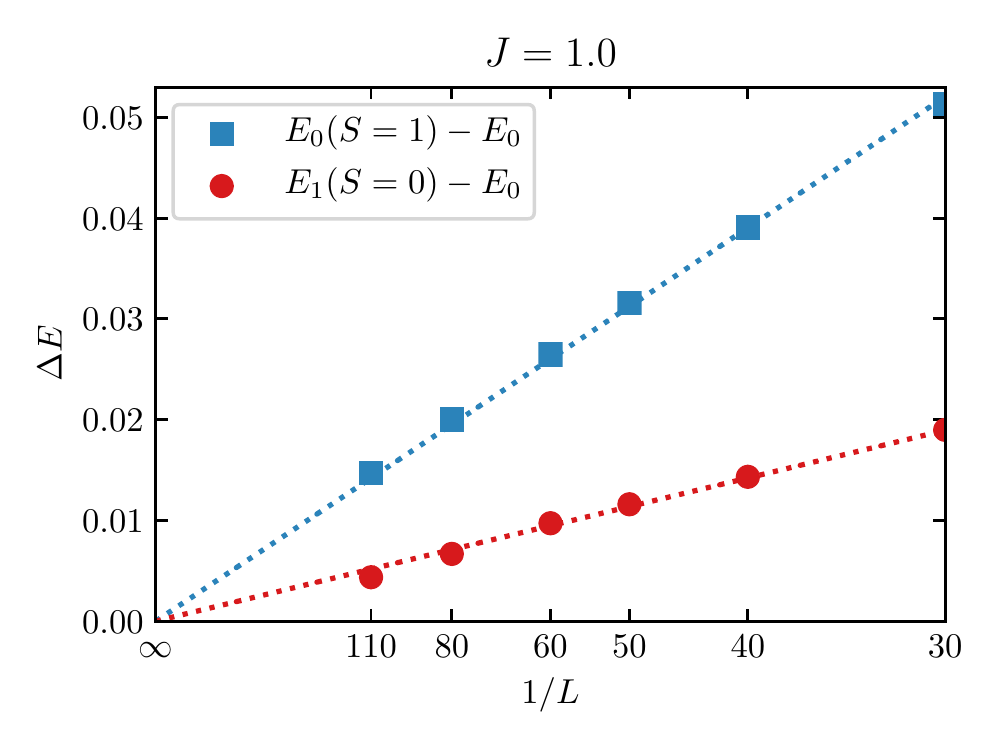}
  \caption{Spin triplet and singlet excitation gaps [$E_0(S=1) - E_0$ and $E_1(S=0) - E_0$, respectively] versus $1/L$ calculated with DMRG on the OBC($<<$) system at $J=1.0$. For generic $L$, the energies/gaps exhibit some nonmonotonic behavior with $1/L$---consistent with finite-size ``shell-filling'' effects for the spinons~\cite{Sheng09_PRB_79_205112, Mishmash11_PRB_84_245127, Jiang13_Nature_493_39} or, relatedly, some system sizes being more compatible with the dominant $q_<$ feature in the bond-energy textures than others---therefore, here we only plot sizes at local minima versus $1/L$. The lines are fits to the simple linear scaling form $\Delta E = a/L$.}
  \label{fig:gaps}
\end{figure}

We conclude this section by presenting additional data on the spin excitation gaps in the putative C1S1 phase. In Fig.~\ref{fig:gaps}, we plot the triplet excitation gap, $E_0(S=1) - E_0$, as well as the singlet excitation gap, $E_1(S=0) - E_0$, versus inverse system length $1/L$ obtained with OBC($<<$) at the characteristic point $J=1.0$. (In the entire interval $0.75\lesssim J\lesssim2.0$, we find that the ground state is a spin singlet with total spin $S=0$; see also Ref.~\cite{Waldtmann00_PRB_62_9472}.)  We show fits to the simple scaling form $\Delta E = a/L$ (not considering log corrections~\cite{Ng96_BetheChainDMRG_PRB_54_9854, Waldtmann00_PRB_62_9472}) to show overall consistency with both gaps vanishing in the thermodynamic limit.  This conclusion is in agreement with previous work~\cite{Waldtmann00_PRB_62_9472, Lauchli}.  Note that the smallest system size ($L=30$) in Fig.~\ref{fig:gaps} is comparable to the largest sizes considered in the early work of Ref.~\cite{Waldtmann00_PRB_62_9472} which also argued for a gapless phase; thus, eventual small spin triplet gaps seem exceedingly unlikely on this kagome strip.

\section{Low-energy bosonized theory for the C1S2 (SBM) and C1S1 states (plus supporting data)} \label{sec:theory}

The long-wavelength description of two gapless 1D bands of spin-1/2 fermions (spinons) coupled to a U(1) gauge field has been treated in detail in Ref.~\cite{Sheng09_PRB_79_205112} (see also Refs.~\cite{Motrunich09_PRB_79_235120, Motrunich10_PRB_81_045105, Ko13_KondoC1S1_PRB_87_205107, Mishmash15_PRB_91_235140}). For brevity, we here only summarize the construction of the theory and highlight those aspects which are most relevant to our results on the kagome strip. Along the way, we will also present some additional numerical data supporting our conclusions in the main text.

\subsection{Bosonization description} \label{sec:bosonization}

We label the two partially filled bands in Fig.~\ref{fig:bands} as $b=a,s$, where band $a$ ($s$) has associated wave functions which are antisymmetric (symmetric) under interchange of the top and bottom legs of the kagome strip. To import results from Ref.~\cite{Sheng09_PRB_79_205112}, we use the band-mapping dictionary $1\leftrightarrow a$ and $2\leftrightarrow s$ and follow the associated bosonization conventions. Taking the low-energy continuum limit, we expand the spinon operators in terms of slowly varying continuum fields $f_{Pb\alpha}$ near the Fermi points~\cite{Giamarchi03_1D}; $P=R/L=+/-$ denotes right and left moving fermion fields, $b=a,s$ is a band index, and $\alpha=\,\uparrow,\downarrow$ is the spin index. At the mean-field level (before introducing gauge fluctuations), we thus have a state with $c=4$ 1D gapless (nonchiral) modes, which in terms of bosonized fields can be expressed as~\cite{Giamarchi03_1D}
\begin{equation}
f_{Pb\alpha} = \eta_{b\alpha}e^{i(\varphi_{b\alpha} + P\theta_{b\alpha})},
\end{equation}
where $\varphi_{b\alpha}$ and $\theta_{b\alpha}$ are the canonically conjugate bosonic phase and phonon fields, respectively, and $\eta_{b\alpha}$ are the Klein factors satisfying Majorana anticommutation relations, $\{\eta_{b\alpha}, \eta_{b'\beta} \} = 2\delta_{bb'}\delta_{\alpha\beta}$~\cite{Sheng09_PRB_79_205112}.  It is natural in this context to take linear combinations of the original four bosonic fields $\theta_{b\alpha}$ which correspond to ``charge'' ($\rho$) and ``spin'' ($\sigma$) modes for each band $b$:
\begin{equation}
\theta_{b\rho/\sigma} = \frac{1}{\sqrt{2}}(\theta_{b\uparrow} \pm \theta_{b\downarrow}),
\end{equation}
as well as ``overall'' and ``relative'' combinations with respect to the two bands:
\begin{equation}
\theta_{\mu\pm} = \frac{1}{\sqrt{2}}(\theta_{a\mu} \pm \theta_{s\mu}),
\end{equation}
where $\mu=\rho,\sigma$. Analogous definitions also hold for the $\varphi$ fields.

Inclusion of gauge fluctuations leads to a mass term for the overall (gauge) charge mode $\theta_{\rho+}$, thus essentially implementing a coarse-grained version of the microscopic on-site constraint $\sum_{\alpha}f_{i\alpha}^\dagger f_{i\alpha} = 1$. From now on we will thus will assume that, up to massive quadratic fluctuations, the field $\theta_{\rho+}$ is pinned. The final resulting state is a two-band analog of the U(1) spinon Fermi surface state (i.e., ``spin Bose metal'' or SBM):~~It is a highly unconventional (insulating) C1S2 Luttinger liquid with one gapless ``relative charge'' mode, $\theta_{\rho-}$, and two gapless spin modes, $\theta_{s\sigma}$ and $\theta_{a\sigma}$ ($c=3$ total 1D gapless modes). The field $\theta_{\rho-}$ has an associated nontrivial Luttinger parameter $g_{\rho-}$, while SU(2) symmetry dictates trivial Luttinger parameters in the spin sector ($g_{a\sigma} = g_{s\sigma} = 1$).  (For the specific quadratic Lagrangian for the SBM fixed point, including relevant bosonization conventions that we employ, we refer the reader to Ref.~\cite{Sheng09_PRB_79_205112}.)

Considering the symmetries present in our kagome strip Heisenberg model---i.e.,
SU(2) spin rotation, time reversal, $x\leftrightarrow-x$ reflection (mirror),
top-bottom leg interchange, and spatial translations along $x$ by one unit
cell---the set of allowed (nonchiral) short-range four-fermion interactions of
the spinons \emph{at generic band-filling configuations} ($k_{Fa}$ and
$k_{Fs}$) are identical to those listed in Ref.~\cite{Sheng09_PRB_79_205112}
(see also Refs.~\cite{Balents96_PRB_53_12133, Lin97_PRB_56_6569,
Motrunich10_PRB_81_045105}).  In terms of the so-called chiral currents,
\begin{equation}
J_{Pbb'} = f^\dagger_{Pb\alpha} f_{Pb'\alpha}\,,~~~\mathbf{J}_{Pbb'} = \frac{1}{2}f^\dagger_{Pb\alpha}\bm{\sigma}_{\alpha\beta} f_{Pb'\beta},
\end{equation}
these interactions can be written as follows:
\begin{align}
\mathcal{H}^\rho_w & = \sum_{b,b'}w_{bb'}^\rho J_{Rbb'}J_{Lbb'}, \\
\mathcal{H}^\rho_\lambda & = \sum_{b,b'}\lambda_{bb'}^\rho J_{Rbb}J_{Lb'b'}, \\
\mathcal{H}^\sigma_w & = -\sum_{b,b'}w_{bb'}^\sigma \mathbf{J}_{Rbb'}\cdot\mathbf{J}_{Lbb'}, \\
\mathcal{H}^\sigma_\lambda & = -\sum_{b,b'}\lambda_{bb'}^\sigma \mathbf{J}_{Rbb}\cdot\mathbf{J}_{Lb'b'},
\end{align}
where $w^{\rho/\sigma}_{aa} = w^{\rho/\sigma}_{ss} = 0$ (convention / absorbed into $\lambda$ terms), $w^{\rho/\sigma}_{as} = w^{\rho/\sigma}_{sa}$ (from Hermiticity), and $\lambda^{\rho/\sigma}_{as} = \lambda^{\rho/\sigma}_{sa}$ (from $R\leftrightarrow L$ symmetry).

A potentially harmful interaction is the so-called $W$ term composed of $\mathcal{H}^\rho_w + \mathcal{H}^\sigma_w$~\cite{Sheng09_PRB_79_205112, Motrunich10_PRB_81_045105}:
\begin{align}
W \equiv & \,(w^\rho_{as}J_{Ras}J_{Las} - w^\sigma_{as}\mathbf{J}_{Ras}\cdot\mathbf{J}_{Las}) + \mathrm{H.c.} \label{eq:W} \\
= &\, \cos(2\varphi_{\rho-})\{4w_{as}^\rho[\cos(2\varphi_{\sigma-}) - \hat\Gamma\cos(2\theta_{\sigma-})] \nonumber \\
- &\, w_{as}^\sigma[\cos(2\varphi_{\sigma-}) + \hat\Gamma\cos(2\theta_{\sigma-}) + 2\hat\Gamma\cos(2\theta_{\sigma+})]\}, \nonumber
\end{align}
where
\begin{align}
\hat\Gamma\equiv\eta_{1\uparrow}\eta_{1\downarrow}\eta_{2\uparrow}\eta_{2\downarrow}.
\end{align}
The $W$ term thus has a scaling dimension of $\Delta[W] = 1 +
\Delta[\cos(2\varphi_{\rho-})] = 1 + \frac{1}{g_{\rho-}}$, and if it is
relevant ($\Delta[W] < 2$), all three gapless modes present in the C1S2 become
gapped leading to some fully gapped C0S0 paramagnet. Hence, stability of the
parent C1S2 state at generic $k_{Fa}, k_{Fs}$ necessarily requires the
condition $g_{\rho-}\leq1$.

Based on the characteristics of the DMRG data in the regime $0.8\lesssim
J\lesssim1.3$, it is natural to explore the situation in which the single-band
$2k_{F}$ backscattering interaction $\lambda_{ss}^\sigma$ is marginally
relevant, while the analogous terms $\lambda_{aa}^\sigma$ and
$\lambda_{as}^\sigma$ are marginally irrelevant. This occurs given that
$\lambda_{ss}^\sigma < 0$, while $\lambda_{aa}^\sigma > 0$ and
$\lambda_{as}^\sigma > 0$~\cite{Balents96_PRB_53_12133, Sheng09_PRB_79_205112}.
We currently have little microscopic intuition for why this might be the case
in our model but proceed based on the scenario's appealing phenomenology. In
terms of bosonized fields, the term $\lambda_{ss}^\sigma$ contains a cosine
potential,
\begin{equation}
V_{ss}^\perp = \lambda_{ss}^\sigma\cos(2\sqrt{2}\theta_{s\sigma}),
\end{equation}
so that relevance of $\lambda_{ss}^\sigma$ pins the field $\theta_{s\sigma}$ associated with the spin mode of band $s$. The resulting state is a C1S1 Luttinger liquid with $c=2$ 1D gapless modes, $\theta_{\rho-}$ and $\theta_{a\sigma}$. We must still require that the $W$ term is irrelevant for C1S1 to be a stable phase. Given that $\theta_{s\sigma}$ is pinned (hence $\varphi_{s\sigma}$ is fluctuating wildly), the important part of the $W$ interaction in terms of bosonized fields reads~\cite{Sheng09_PRB_79_205112}
\begin{equation}
W = -(4w^\rho_{as} + 3w^\sigma_{as})\cos(\sqrt{2}\theta_{a\sigma})\cos(\sqrt{2}\theta_{s\sigma})\cos(2\varphi_{\rho-}),
\label{eq:WC1S1}
\end{equation}
where now $\theta_{s\sigma}$ is pinned, while $\theta_{a\sigma}$ and
$\varphi_{\rho-}$ are both fluctuating. The scaling dimension of the $W$ term
with respect to the C1S1 fixed point is thus $\Delta[W] = \frac{1}{2} +
\frac{1}{g_{\rho-}}$, so that stability of the C1S1 state at generic $k_{Fa}$,
$k_{Fs}$ further requires $g_{\rho-} < 2/3$.

\subsection{Observables} \label{sec:observables}

To connect to the DMRG measurements of bond-energy textures and spin-spin
correlations functions, we now turn to bosonized expressions of the bond-energy
and spin operators at finite wave vectors. We first consider fermion bilinears
and focus on those composed of a (spinon) particle and hole moving in opposite
directions, i.e., Amperean-enhanced contributions~\cite{Lee06_RevModPhys_78_17,
Sheng09_PRB_79_205112}. For spin operators symmetric under leg interchange,
e.g., $\mathbf{S}_x^{s}$ and $\mathbf{S}_x^{M}$, by symmetry we can write down
the following contributions at wave vectors $2k_{Fb}$:
\begin{align}
\mathbf{S}_{2k_{Fb}} & = \frac{1}{2}f^\dagger_{Lb\alpha}\bm{\sigma}_{\alpha\beta}f_{Rb\beta}, \label{eq:S} \\
S^x_{2k_{Fb}} & \propto e^{i\theta_{\rho+}}e^{\pm i\theta_{\rho-}}\sin(\sqrt{2}\varphi_{b\sigma}), \label{eq:Sx} \\
S^y_{2k_{Fb}} & \propto e^{i\theta_{\rho+}}e^{\pm i\theta_{\rho-}}\cos(\sqrt{2}\varphi_{b\sigma}), \label{eq:Sy} \\
S^z_{2k_{Fb}} & \propto e^{i\theta_{\rho+}}e^{\pm i\theta_{\rho-}}\sin(\sqrt{2}\theta_{b\sigma}) \label{eq:Sz},
\end{align}
while for the bond energy at $2k_{Fb}$, we have
\begin{align}
\varepsilon_{2k_{Fb}} & = \frac{1}{2}f^\dagger_{Lb\alpha}f_{Rb\alpha} \label{eq:fermBq}, \\
\mathcal{B}_{2k_{Fb}} \propto \varepsilon_{2k_{Fb}} & \propto e^{i\theta_{\rho+}}e^{\pm i\theta_{\rho-}}\cos(\sqrt{2}\theta_{b\sigma}). \label{eq:bosBq}
\end{align}
(In these expressions, $\pm$ corresponds to band $b=a/s$.) Note that at the C1S2 and C1S1 fixed points, the overall charge mode is pinned in the above expressions, i.e., $\theta_{\rho+} = \mathrm{const.}$

On the other hand, for the spin operator $\mathbf{S}_x^{a}$, which is
antisymmetric under leg interchange, we have analogous contributions at wave
vector $\pi/2$. In addition, the bottom-leg bond-energy texture $\mathcal{B}_q$
defined above, which has no simple transformation property under leg
interchange, would also have a contribution at $\pi/2$. (We refer the reader to
Ref.~\cite{Sheng09_PRB_79_205112} for the detailed expressions in each case.)

\begin{table}[t!]
\begin{tabular}{| c || c || c | c || c | c |}
\hline
 & \,$c$\, & $\Delta[\mathcal{B}_{2k_{Fs}}]$ & $\Delta[\mathcal{B}_{2k_{Fa}}]$ & $\Delta[\mathbf{S}_{2k_{Fs}}]$ & $\Delta[\mathbf{S}_{2k_{Fa}}]$  \\
\hline
\hline
  \begin{tabular}{@{}c@{}}C1S2 \\ (SBM)\end{tabular} & \,3\, & $\frac{1}{2} + \frac{g_{\rho-}}{4}$ & $\frac{1}{2} + \frac{g_{\rho-}}{4}$ & $\frac{1}{2} + \frac{g_{\rho-}}{4}$  & $\frac{1}{2} + \frac{g_{\rho-}}{4}$ \\
\hline
  \begin{tabular}{@{}c@{}}C1S1 \\ (realized)\end{tabular} & \,2\, & $\frac{g_{\rho-}}{4}$ & $\frac{1}{2} + \frac{g_{\rho-}}{4}$ & $\infty$ & $\frac{1}{2} + \frac{g_{\rho-}}{4}$ \\
\hline
  \begin{tabular}{@{}c@{}}C0S1 \\ (BCS wf)\end{tabular} & \,1\, & $\infty$ & $\frac{1}{2}$ & $\infty$ & $\frac{1}{2}$ \\
\hline
\end{tabular}
\caption{Central charge, $c$, and scaling dimensions of the bond-energy and spin operators at wave vectors $q_< = 2k_{Fs}$ and $q_> = 2k_{Fa}$ for the C1S2, C1S1, and C0S1 states.  C1S2 is the SBM theory whose wave functions we compare directly with the DMRG.  C1S1 is the phase which we argue is actually realized in the DMRG.  Finally, C0S1 refers to the BCS wave function described below in Appendix~\ref{sec:vmc} which would (relative to the DMRG) correctly capture short-ranged ($\Delta=\infty$) spin correlations at wave vector $q_< = 2k_{Fs}$, but it would also incorrectly (and tragically) give rise to short-ranged bond-energy correlations at wave vector $q_< = 2k_{Fs}$ as well as central charge $c=1<2$, both of which are qualitatively inconsistent with C1S1 and the DMRG.  The dominant feature in the C1S1 phase is in fact that in the bond energy at $q_< = 2k_{Fs}$; cf.~the DMRG data in Fig.~\ref{fig:bondtextures}.}
\label{tab:2kFops}
\end{table}

From the above discussion, it is clear that in the C1S2 (SBM) state we should
in general expect power-law singularities in
$\langle\mathbf{S}_{q}^{s}\cdot\mathbf{S}_{-q}^{s}\rangle$ and
$\langle\mathbf{S}_{q}^{M}\cdot\mathbf{S}_{-q}^{M}\rangle$ at wave vectors $q_<
= 2k_{Fs}$ and $q_> = 2k_{Fa}$ and similarly in
$\langle\mathbf{S}_{q}^{a}\cdot\mathbf{S}_{-q}^{a}\rangle$ at wave vector
$\pi/2$. This is fully consistent with our VMC calculations, as shown, for
example, in Fig.~\ref{fig:ssf}. The structure factor
$\langle\mathbf{S}_{q}^{B}\cdot\mathbf{S}_{-q}^{M}\rangle$ could in principle
have contributions at all three wave vectors $2k_{Fs}, 2k_{Fa}$, and $\pi/2$
(although the VMC measurements only show the first two). The same expectations
arise for the Fourier transform of the bond-energy textures
$\mathcal{B}_q^\mathrm{leg}$ and $\mathcal{B}_q^\mathrm{cross}$ (see, for
example, Eqs.~\eqref{eq:fermBq}--\eqref{eq:bosBq} and
Ref.~\cite{Motrunich09_PRB_79_235120}). As displayed in
Fig.~\ref{fig:bondtextures}, the VMC clearly shows features in
$\mathcal{B}_q$ at $q_< = 2k_{Fs}$ and $q_> = 2k_{Fa}$.

If the term $\lambda_{ss}^\sigma$ is relevant---as is putatively realized in
the DMRG state---then subsequent pinning of $\theta_{s\sigma}$ will affect
physical operators such as the spin and bond energy in a qualitative way. By
Eqs.~\eqref{eq:Sx}--\eqref{eq:Sz}, one obvious effect is to eliminate the
power-law feature in the structure factors
$\langle\mathbf{S}_{q}^{s}\cdot\mathbf{S}_{-q}^{s}\rangle$ and
$\langle\mathbf{S}_{q}^{M}\cdot\mathbf{S}_{-q}^{M}\rangle$ at wave vector $q_<
= 2k_{Fs}$. All features at $q=\pi/2$ in both
$\langle\mathbf{S}_{q}^{a}\cdot\mathbf{S}_{-q}^{a}\rangle$ and $\mathcal{B}_q$
are similarly eliminated. (In all these cases, the operator in question
contains the wildly fluctuating field $\varphi_{s\sigma}$, thus leading to
exponential decay in real space.) On the other hand, as can be inferred from
Eq.~\eqref{eq:bosBq}, the bond energy at wave vector $q_< = 2k_{Fs}$ actually
gets \emph{enhanced} upon pinning of $\theta_{s\sigma}$, i.e., slower decay in
real space with concomitant stronger feature in momentum space. We summarize
these points in Table~\ref{tab:2kFops} where we list the scaling dimensions of
the $2k_{F}$ contributions to the bond-energy and spin operators with respect
to both the C1S2 (SBM) and C1S1 fixed points. All in all, a C1S1 state obtained
by (marginal) relevance of $\lambda_{ss}^\sigma$ would qualitatively agree with
all features observed in the DMRG data in Figs.~\ref{fig:bondtextures} and
\ref{fig:ssf}. Unfortunately, as we discuss below in Appendix~\ref{sec:vmc},
faithfully representing such a C1S1 state with a Gutzwiller-projected
variational wave function cannot be accomplished in a straightforward way.

\begin{figure}[t]
  \centering
  \includegraphics[width=0.95\columnwidth]{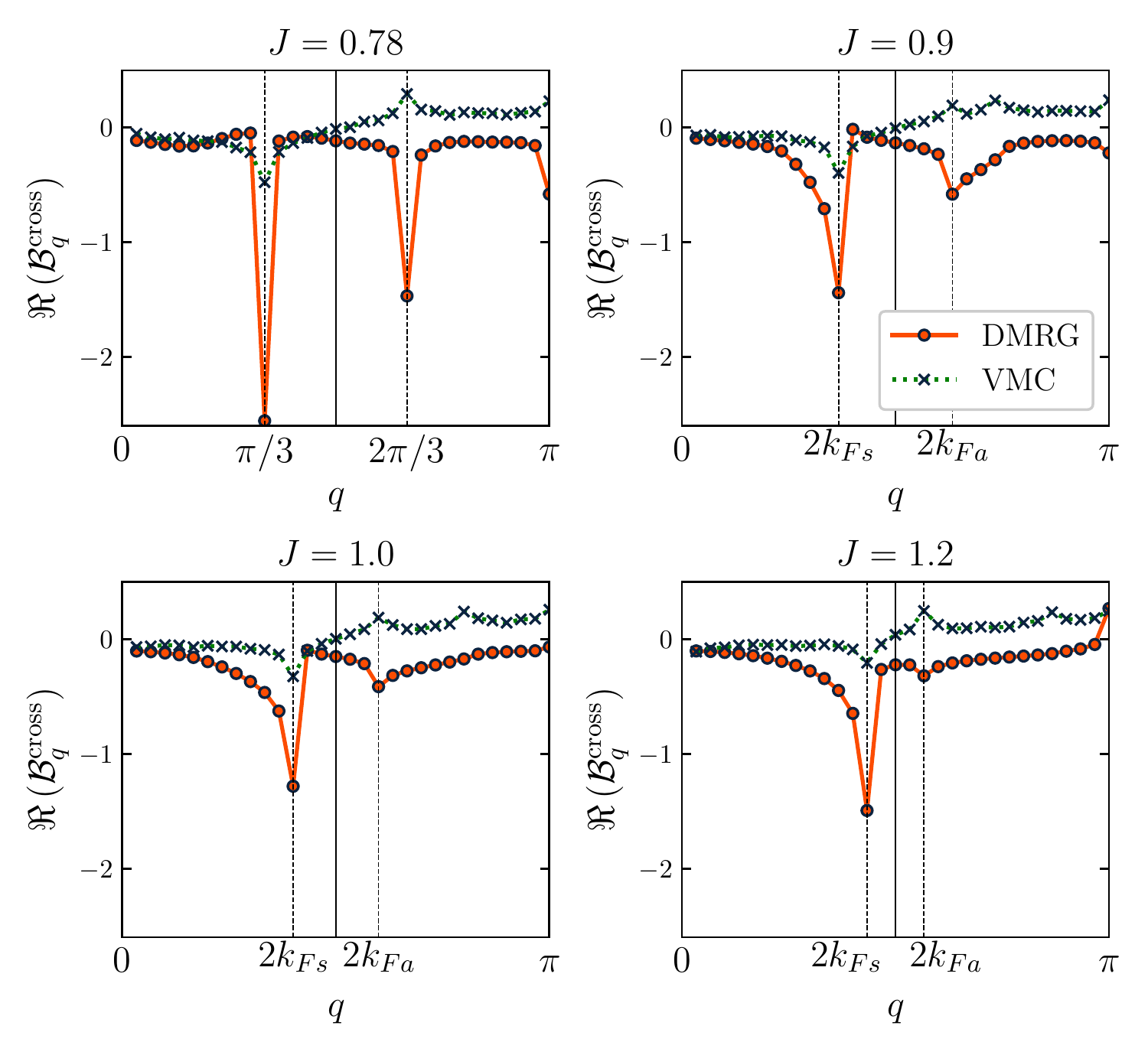}
  \caption{Data analogous to Fig.~\ref{fig:bondtextures}, but now taking the Fourier transform of the \emph{cross-bond} bond-energy textures [see Eq.~\eqref{eq:Bqcrossdef}].  The parameters chosen for the VMC states in these calculations (and in the analogous calculations of $\mathcal{B}_{q} \equiv \mathcal{B}_{q}^\mathrm{leg}$ in Fig.~\ref{fig:bondtextures}) are $t_{c}=1.0$ and $\mu=-1.8,-2.4,-3.1,-4.8$ for $J=0.78,0.9,1.0,1.2$, respectively.  (For details of our VMC calculations, please see Appendix~\ref{sec:vmc}.)  Recall that the DMRG ground state at $J=0.78$ is a period-6 VBS (C0S0), but we still show a corresponding VMC state (C1S2) for comparison.  The discrepancies in signs of the features at, for example, $q_> = 2k_{Fa}$ between the DMRG and VMC results can plausibly be explained by nonuniversal amplitudes/phases of the bond texture's oscillatory components (see text). }
  \label{fig:crossbonds}
\end{figure}

In addition, we note that there are potential \emph{four-fermion} contributions
to the spin operator at wave vector $\pi$ and to the bond energy at wave
vectors $4k_{Fa} = -4k_{Fs}$ and $\pi$~\cite{Sheng09_PRB_79_205112} (these
basically arise from \emph{two} $2k_{F}$ processes). For the spin correlations
at $q=\pi$ (see Fig.~\ref{fig:ssf}), there are no such
features in either the DMRG data nor VMC data except for the ``bottom-middle''
structure factor $\langle\mathbf{S}_{q}^{B}\cdot\mathbf{S}_{-q}^{M}\rangle$,
where both the DMRG and VMC show a possible singularity. Turning to the
bond-energy textures, we see in Fig.~\ref{fig:bondtextures}
that neither the DMRG data nor the VMC data possess any obviously noticeable
features at $q=4k_{Fa}$ nor at $q=\pi$ in $\mathcal{B}_q^\mathrm{leg}$
(although the DMRG may indeed show a weaker feature at $4k_{Fa}$). By a scaling
dimension analysis alone, singular structure at $4k_{Fa}$ may be expected to be
comparable to that at $q_>=2k_{Fa}$\,: the scaling dimensions of the bond
energy at the two wave vectors are $g_{\rho-}$ and $\frac{1}{2} +
\frac{g_{\rho-}}{4}$, respectively, with $g_{\rho-} < 2/3$ required for a
stable C1S1. However, nonuniversal amplitudes---which are impossible to predict
with the bosonized gauge theory---also strongly dictate the visibility of
a state's power-law singularities. Such effects are likely to be at play here
in describing, for example, why the VMC state itself shows no singular
structure at $q=4k_{Fa}$ in $\mathcal{B}_q^\mathrm{leg}$ (and similarly for the
DMRG).

In Fig.~\ref{fig:crossbonds}, we present data on cross-bond bond-energy
textures $\mathcal{B}_q^\mathrm{cross}$ [see Eq.~\eqref{eq:Bqcrossdef}]. This
data is analogous to the $\mathcal{B}_q^\mathrm{leg}$ data of
Fig.~\ref{fig:bondtextures}, and it was also obtained with
OBC($<<$). In this case, the VMC data does exhibit features at $q=4k_{Fa}$ and
$q=\pi$, while the DMRG clearly shows a feature only at $q=\pi$. (Although, as
in $\mathcal{B}_q^\mathrm{leg}$, the DMRG data may have a weak feature at
$4k_{Fa}$ if one looks closely---the fact that it is not stronger is plausibly
due to the amplitude effect described above). Note that the features at $q_> =
2k_{Fa}$ have opposite signs in the DMRG and VMC data sets. However, the
amplitudes and phases of these bond-energy textures are known to be
nonuniversal and strongly dependent on the details of the pinning conditions at
the boundary~\cite{Motrunich09_PRB_79_235120}. For our VMC calculations with
open boundaries, we form a Gutzwiller-projected Fermi sea wave function
obtained by simply diagonalizing a free spinon hopping Hamiltonian with uniform
hopping amplitudes along the $x$ direction (see Appendix~\ref{sec:vmc} below) but with
hard-wall boundary conditions. We have attempted tuning the details of this
hopping Hamiltonian (e.g., magnitudes and signs of the hopping amplitudes) near
the boundary with the hope of flipping the sign of the $q_> = 2k_{Fa}$ feature
in $\mathcal{B}_q^\mathrm{cross}$. Although by doing so we were able to
drastically alter the magnitudes of the features, we were unsuccessful in
flipping the sign of the $q_> = 2k_{Fa}$ feature. Still, this should be
possible in principle. As an explicit example of how the signs of such singular
features are nonuniversal, we would like to point out the following observation
about the behavior at $q=\pi$ in Fig.~\ref{fig:crossbonds}: In the DMRG data
itself, the feature at $q=\pi$ actually appears to flip sign as one tunes
through the phase from $J=0.9$ (where the feature has ``negative'' sign) to
$J=1.2$ (where it has ``positive'' sign).

\begin{figure}[t]
  \centering
  \includegraphics[width=0.95\columnwidth]{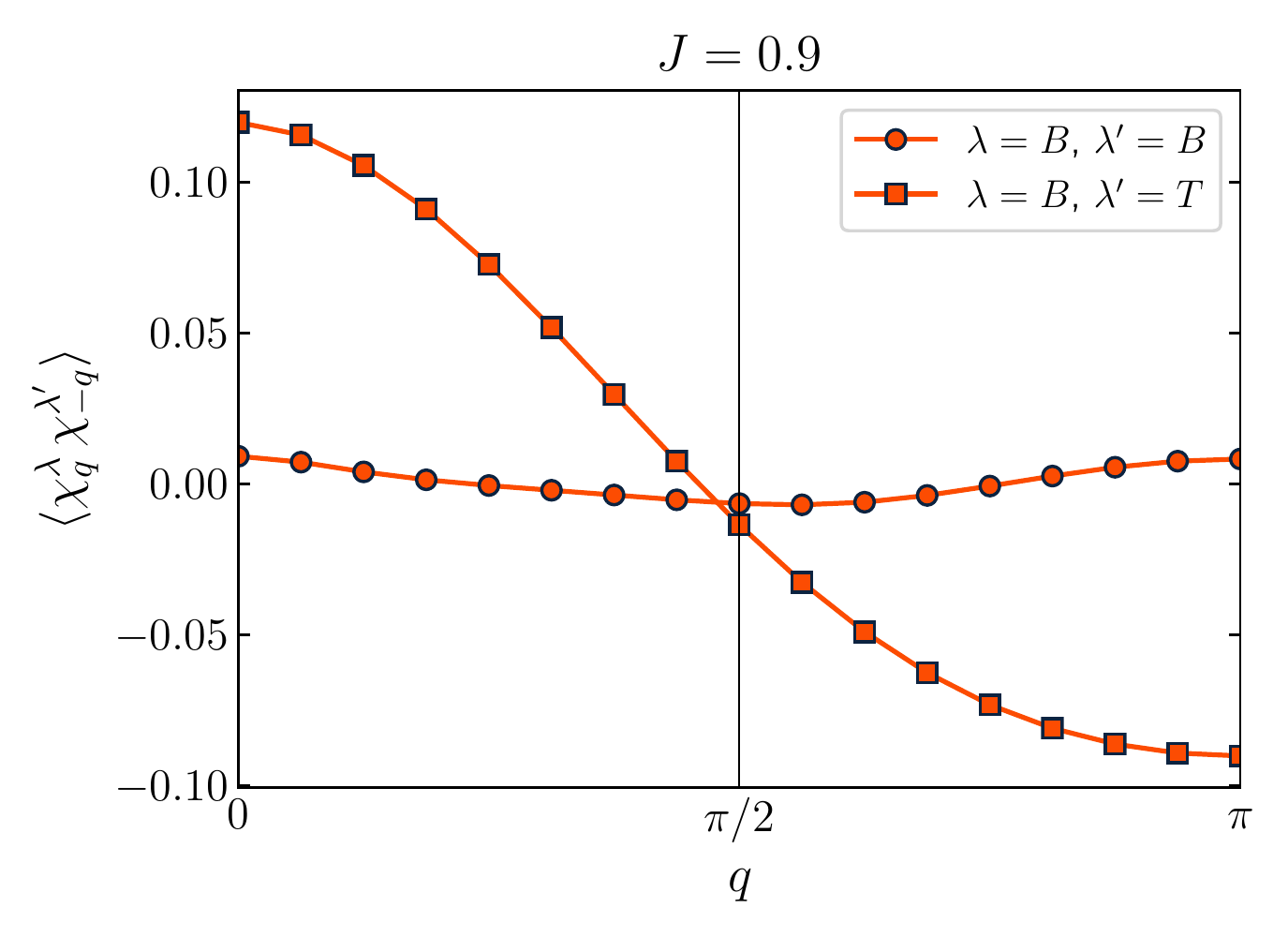}
  \caption{Chirality structure factors obtained with DMRG at the point $J=0.9$ on a PBC system of length $L=32$; the specific quantities being plotted are detailed in Eqs.~\eqref{eq:chichiBB}--\eqref{eq:chix}. The lack of Bragg peaks implies that the DMRG ground state respects time-reversal symmetry (which we have also verified with complex-valued DMRG simulations); and the lack of power-law singularities at finite wave vectors indicates short-ranged behavior at those wave vectors in the chirality sector. This behavior is consistent with the proposed C1S1 theory.}
  \label{fig:chichi}
\end{figure}

As a final characterization of the DMRG ground state in the regime $0.8\lesssim J\lesssim1.3$, we present in Fig.~\ref{fig:chichi} measurements of the chirality structure factors defined in Eqs.~\eqref{eq:chichiBB}--\eqref{eq:chichiBT} at the representative point $J=0.9$. We see that these Fourier-space measurements (1) are featureless at finite wave vectors and (2) exhibit no Bragg peaks. Both of these properties are predicted by the C1S1 theory: (1) Gapping of the spin mode $\theta_{s\sigma}$ will result in short-ranged decay of the chirality-chirality correlations at all finite wave vectors (see discussion in the main text and Appendix A of Ref.~\cite{Sheng09_PRB_79_205112}), and (2) the theory respects time-reversal symmetry. Note, however, that the $\rho-$ part of the theory can still produce $1/x^2$ decay at zero momentum with nonuniversal prefactors~\cite{Sheng09_PRB_79_205112}. There are noticeable corresponding slope discontinuities at $q=0$ in the data in Fig.~\ref{fig:chichi}---we believe the relatively small slopes are merely a quantitative matter. In fact there are similarly weak $q=0$ slope discontinuities in the spin structure factor measurements (even in some of the VMC data), while we know with absolute certainty that the spin sector is gapless; furthermore, weak slope discontinuities in $\langle\chi_q\chi_{-q}\rangle$ at $q=0$ were likewise observed in the C1S2 SBM phase of Ref.~\cite{Sheng09_PRB_79_205112} (see e.g.~their Fig.~5). All in all, the chirality structure factors exhibited by the DMRG are fully consistent with the universal properties of the spin chirality sector of the C1S1 phase.

\subsection{Instabilities out of C1S1}

In this section, we describe the situation for the states peripheral to the region $0.8\lesssim J\lesssim1.3$.  Notably, the instability for $J\lesssim0.8$ can be very naturally described within the C1S1 theory, while that for $J\gtrsim1.3$ occurs via a strong first-order phase transition---possibly even intervening phase---and likely lies outside of our theoretical framework (but see below).

In the DMRG, we observe a state with long-range (dominant) period-6 VBS order (VBS-6) for $0.75\lesssim J\lesssim0.8$.  Tracking the singular wave vectors in the DMRG, we expect this state to correspond to $q_<=2k_{Fs} = \pi/3, q_>=2k_{Fa} = 2\pi/3$ ($k_{Fs}=5\pi/6, k_{Fa}=2\pi/3$).  (Such equalities involving wave vectors are implied to mean so up to signs and mod\,$2\pi$.)  Indeed, when the theory is at the special commensurate point corresponding to $k_{Fs} = 5\pi/6$ and $k_{Fa} = 2\pi/3$, there is an additional symmetry-allowed six-fermion umklapp-type interaction which needs to be considered:
\begin{align}
V_6 & = u_6(f^\dagger_{Rs\uparrow}f^\dagger_{Rs\downarrow}f^\dagger_{La\alpha}f_{Ls\uparrow}f_{Ls\downarrow}f_{Ra\alpha} + \mathrm{H.c.}) \\
& = -4u_6 \cos(\sqrt{2}\theta_{a\sigma})\sin(3\theta_{\rho-} - \theta_{\rho+}).
\end{align}
This term has scaling dimension with respect to the C1S1 (and C1S2) fixed point of $\Delta[V_6] = \frac{1}{2}+\frac{9}{4}g_{\rho-}$ and is thus relevant given $g_{\rho-}< 2/3$.  Since this is precisely the condition required for the $W$ term to be irrelevant and thus C1S1 to be a stable phase at generic $k_{Fs}$ and $k_{Fa}$, a C1S1 state tuned to the point $k_{Fs} = 5\pi/6$ and $k_{Fa} = 2\pi/3$ must necessarily be unstable to this interaction.  Relevance of $V_6$ thus pins both of the remaining gapless modes, $\theta_{a\sigma}$ and $\theta_{\rho-}$, in the C1S1 phase. Inspection of Eq.~\eqref{eq:bosBq} reveals that the resulting fully gapped C0S0 state would have coexisting period-6 and period-3 VBS order (with the former being dominant).

\begin{figure}
  \centering
  \includegraphics[width=0.95\columnwidth]{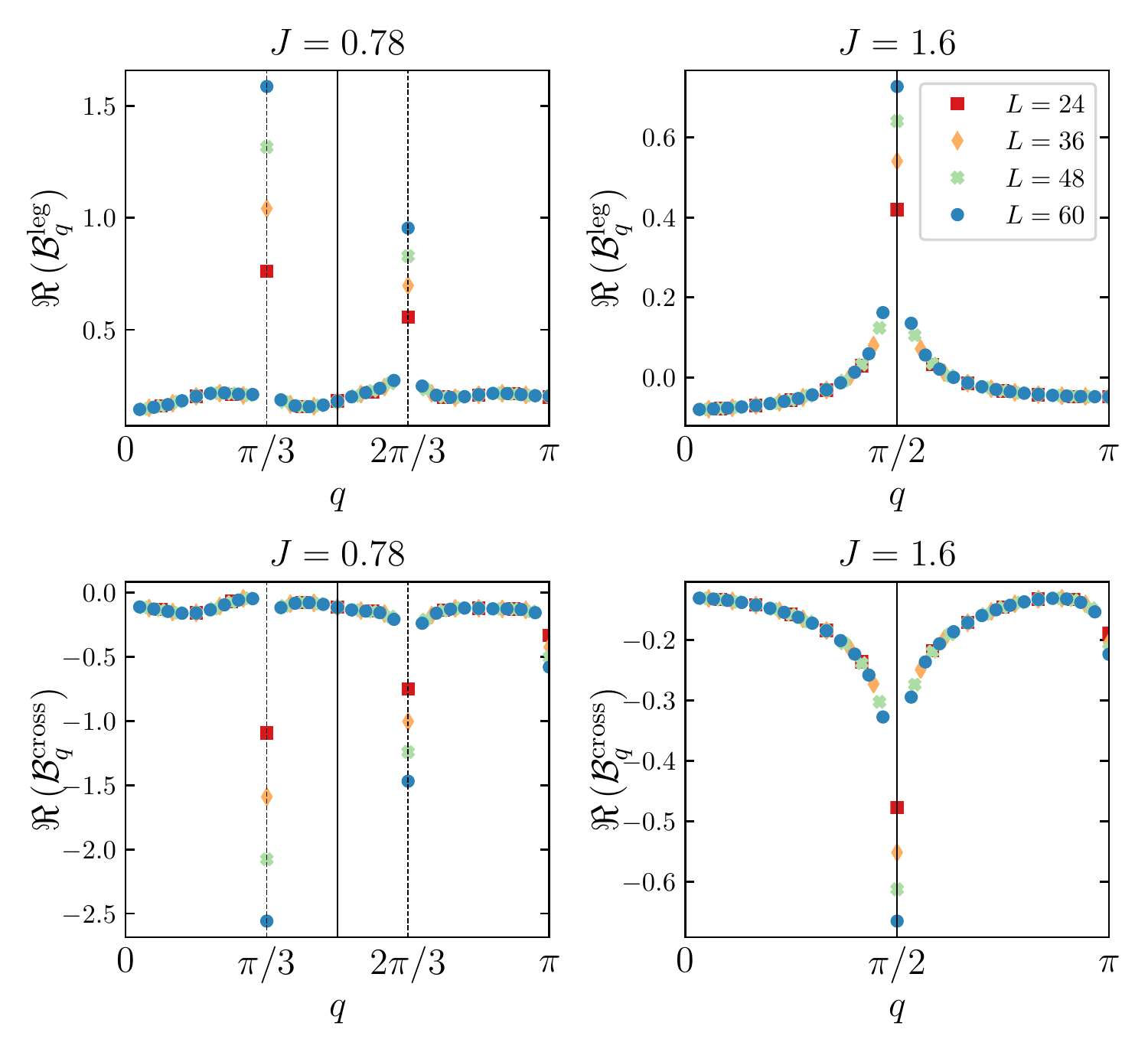}
  \caption{Leg-bond (top row) and cross-bond (bottom row) bond-energy texture data for a sequence of lengths $L$ in the period-6 VBS phase at $J=0.78$ (left column) and in the period-4 phase at $J=1.6$ (right column).  Development of Bragg peaks in the former case is evident.  These calculations were performed with DMRG using the OBC($<<$) geometry.}
  \label{fig:VBSs}
\end{figure}

As remarked above, we would anticipate this state to be realized in the kagome strip Heisenberg model for values of $J$ just below 0.8. Remarkably, we indeed find evidence for a state with long-range period-6 and period-3 VBS order in the narrow region $0.75\lesssim J\lesssim0.8$.  In Fig.~\ref{fig:VBSs}, we show bond-energy texture data ($\mathcal{B}_q^\mathrm{leg/cross}$) taken with DMRG at a characteristic point $J=0.78$ within this narrow window for a sequence of system sizes on the OBC($<<$) geometry.  We see clear development of Bragg peaks at wave vectors $q=2\pi/6$ and $q=2\pi/3$ in both $\mathcal{B}_q^\mathrm{leg}$ and $\mathcal{B}_q^\mathrm{cross}$ as advertised.  (We also see a potential Bragg peak at wave vector $q=\pi$ in $\mathcal{B}_q^\mathrm{cross}$---as discussed above, such period-2 activity also naturally arises from the theory~\cite{Sheng09_PRB_79_205112}.)  Convergence of the DMRG in this region of the phase diagram is challenging, and we have thus not been able to conclusively determine that the system is fully gapped (e.g., through explicit spin gap calculations, spin-spin correlation functions, or entanglement entropy measurements), although indications are that it likely is (also consistent with Ref.~\cite{Lauchli}).  Near $J\simeq0.75$, a first-order phase transition occurs, and for $J\lesssim0.75$, it appears our theory based on two bands of fermionic spinons no longer applies.  We experience strange convergence difficulties in the DMRG for $0.5\lesssim J\lesssim0.75$, and we have not thoroughly examined the situation for $J\lesssim0.5$.  In fact, it is even an interesting open question whether or not the decoupled Bethe chain phase at $J=0$ persists to any finite $J$.

Next we discuss the behavior for $J\gtrsim1.3$.  For $1.3\lesssim J\lesssim1.4$, the system exhibits strange behavior (and DMRG convergence difficulties) consistent with a strong-first order phase transition, while for $1.4\lesssim J\lesssim2.0$ the DMRG state displays (likely) power-law decaying bond-energy textures with period-4.  There does exist an additional four-fermion momentum-conserving interaction at the special point of the theory when $k_{Fs} = k_{Fa} = 3\pi/4$.  [This term is closely analogous to the $W$ term in Eq.~\eqref{eq:W}---the two have equivalent operator forms upon taking $a\leftrightarrow s$ in the band indices for $J_{Lbb'}$ and $\mathbf{J}_{Lbb'}$.]  One can show that this interaction has scaling dimensions with respect to the C1S2 and C1S1 fixed points of $1+g_{\rho-}$ and $\frac{1}{2}+g_{\rho-}$, respectively, and is thus always relevant if the generic states are stable (i.e., if $W$ is irrelevant).  The resulting state is a some fully gapped C0S0 paramagnet with long-range period-4 VBS order.  This is not consistent with the DMRG data for $1.4\lesssim J\lesssim2.0$ which is (likely) gapless (see also Refs.~\cite{Waldtmann00_PRB_62_9472, Lauchli}) with power-law decaying bond-energy correlations (however, Ref.~\cite{Lauchli} does report a finite VBS-4 order parameter, and we cannot rule out eventual small gaps).  In Fig.~\ref{fig:VBSs}, we show bond-energy texture data for $J=1.6$, which is representative of the behavior in this period-4 phase.  Again, since this state is entered through a strong first-order phase transition near $J\simeq1.3$ (the DMRG exhibits convergence difficulties for $J\simeq1.3-14$), it is thus not surprising that the realized period-4 phase is not naturally accessible starting from the C1S1 theory.
Finally, for $J\gtrsim2.0$, the ground state is a conventional quasi-1D ferrimagnet continuously connected to that realized for $J\to\infty$~\cite{Waldtmann00_PRB_62_9472}.

We conclude by remarking that the bond-energy textures in the putative C1S1 phase itself ($0.8\lesssim J\lesssim1.3$) definitively exhibit power-law decay; this can be gleaned from the Fourier-space data in Figs.~\ref{fig:bondtextures} and \ref{fig:crossbonds}, and we have also performed a complementary real-space analysis.  Within this phase, there is no VBS ordering tendency: For example, the $L=60$ system would be able to accommodate potential VBS states with periods 4, 5, or 6, but for $0.8\lesssim J\lesssim1.3$ the singular wave vectors are incommensurate and fully tunable.

\subsection{Comparison to results of Azaria et al. (Ref.~\cite{azaria_chiral_1998})} \label{sec:azaria}

Reference~\cite{azaria_chiral_1998} described a $c=2$ fixed point in a frustrated three-leg spin ladder, and it is natural to explore the relationship between this fixed point and our C1S1 phase. Ultimately, however, our C1S1 state cannot be accessed in any meaningful way from the ladder model discussed in Ref.~\cite{azaria_chiral_1998}. Firstly, the fixed point at the focus of Ref.~\cite{azaria_chiral_1998} is accessed \emph{perturbatively} via weakly coupling three Heisenberg (Bethe) chains. This is in sharp contrast to our results in which the underlying lattice does not consist of three decoupled chains in any limit; more generally, our C1S1 theory clearly cannot be accessed via weakly coupled chains---one needs to start with incommensurate filling of multiple fermionic spinon bands. The fixed point of Ref.~\cite{azaria_chiral_1998}, in contrast to the C1S1 phase, exhibits only commensurate correlations. While it is in principle possible to reach a phase with incommensurate wavevectors starting from decoupled Heisenberg chains, in general such approaches require terms that manifestly break the SU(2) symmetry of the Hamiltonian~\cite{nersesyan_incommensurate_1998}. The intriguing point about our results is the observation that a simple nearest-neighbor Heisenberg Hamiltonian that retains SU(2) symmetry harbors a phase at low energies with incommensurate wavevectors obeying ``Fermi-like'' sum rules. Furthermore, our C1S1 state is observed over an extended region of parameter space and is thus a \emph{stable quantum phase}. This means that all short-range interactions that are allowed by symmetry are either irrelevant or marginally irrelevant. This is markedly different from an \emph{unstable fixed point} such as the one discussed in Ref.~\cite{azaria_chiral_1998}, where gapless behavior requires relevant perturbations be fine-tuned to zero.


\section{Details of the VMC calculations and projected wave functions} \label{sec:vmc}

For our variational Monte Carlo calculations, we construct a given trial wave function in the standard way by projecting out doubly-occupied sites (``Gutzwiller projection'') from the ground state of a free-fermion (mean-field) Hamiltonian.  In the case of the SBM, this procedure is particularly simple as the mean-field Hamiltonian is a pure hopping model~\cite{Sheng09_PRB_79_205112, Motrunich05_PRB_72_045105}:
\begin{equation}
H_\mathrm{MF} = -\sum_{i,j} t_{ij} f_{i\alpha}^\dagger f_{j\alpha}.
\end{equation}
Here, the sum over spin indices $\alpha=\,\uparrow,\downarrow$ is implied, Hermiticity requires $t_{ij} = t_{ji}^*$, and the on-site ``chemical potential'' terms are given by the diagonal elements: $t_{ii}\equiv\mu_i$.  We then diagonalize $H_\mathrm{MF}$, construct a spin-singlet free-fermion Slater determinant $|\Psi_0(\{t_{ij}\})\rangle$ at half filling from the $N_\uparrow = N_\downarrow = N_\mathrm{sites}/2$ lowest-energy single-particle eigenstates of $H_\mathrm{MF}$, and finally Gutzwiller project:
\begin{equation}
|\Psi_\mathrm{SBM}(\{t_{ij}\})\rangle = \mathcal{P}_G |\Psi_0(\{t_{ij}\})\rangle.
\label{eq:HMFhop}
\end{equation}
The set of hopping amplitudes $\{t_{ij}\}$ defining $H_\mathrm{MF}$ thus constitute the variational parameters of SBM trial states.  These are the ``bare'' Gutwiller states referred to in the main text.  They can be sampled efficiently using standard VMC techniques~\cite{Gros89_AnnPhys_189_53, Ceperley77_PRB_16_3081}.

On the kagome strip, we take hopping strengths of $t_\ell=1$ for the nearest-neighbor leg bonds (orange bonds in Fig.~\ref{fig:clusters}) and $t_{c}\in\mathbb{R}$ for the nearest-neighbor cross bonds (blue bonds in Fig.~\ref{fig:clusters}). Our choice of real values for $t_\ell$ and $t_c$ is justified by the lack of time-reversal symmetry breaking in the DMRG ground state (see Fig.~\ref{fig:chichi}).  Since we are filling up the Fermi sea ``by hand'' the overall chemical potential in $H_\mathrm{MF}$ is arbitrary.  However, still maintaining leg-interchange symmetry between the top and bottom legs, we can have different chemical potentials for the sites on the outer legs (``top'' and ``bottom'') and the ``middle'' chain; we set the former to zero and the latter to $\mu$.  The ansatz thus contains two (real) variational parameters: $t_{c}$ and $\mu$.  For a translationally invariant system, we have a three-site unit cell and $H_\mathrm{MF}$ can be diagonalized analytically resulting in the following band energies as functions of momentum $k$ along the $x$ direction:
\begin{gather}
\epsilon_a(k) = -2t_\ell\cos(k), \\
\epsilon_s^\pm(k) = -\frac{1}{2}\left(\mu - \epsilon_a(k) \mp \sqrt{[\mu + \epsilon_a(k)]^2 + 16t_{c}^2[1+\cos(k)]}\right).
\end{gather}
These bands are shown in Fig.~\ref{fig:bands}, where there we denote the bottommost band $\epsilon_s^-(k)\equiv\epsilon_s(k)$.  (We also show these dispersions again below in Fig.~\ref{fig:VMCstate}, where we discuss the precise state VMC state used in Fig.~\ref{fig:ssf}.)  The corresponding wave functions (with the basis states ordered as ``top'', ``middle'', ``bottom'') are given by
\begin{gather}
\psi_a(k) = \frac{1}{\sqrt{2}}\begin{pmatrix}1 \\ 0 \\ -1\end{pmatrix}, \label{eq:anti} \\
\psi_s^\pm(k) = \frac{1}{\sqrt{2+|\alpha_\pm(k)|^2}}\begin{pmatrix}1 \\ \alpha_\pm(k) \\ 1\end{pmatrix}, \label{eq:sym}
\end{gather}
where $\alpha_\pm(k) \equiv [\epsilon_a(k) -
\epsilon_s^\pm(k)]/[t_{c}(1+e^{ik})]$.
Therefore, band ``$a$'' is antisymmetric under leg interchange, while both
bands ``$s$'' are symmetric.  At $\mu=0$, the bottommost (symmetric) band is
completely filled, while the middle (antisymmetric) band is exactly half
filled; this state does not give rise to the incommensurate structure observed
in the DMRG.  Hence, we focus on the regime $\mu<0$ which produces two
partially filled 1D bands (see Fig.~\ref{fig:bands} and Fig.~\ref{fig:VMCstate} below).
(For discussion of our VMC setup with open boundary conditions, please see Appendix~\ref{sec:observables} above.)

The SBM states described above are model wave functions for the C1S2 phase, while all along we have argued for a C1S1 state as the ground state of the kagome strip Heisenberg model.  A natural question thus concerns how to faithfully described the C1S1 phase via variational wave functions.  Unfortunately, this appears to be nontrivial within the standard paradigm of constructing trial states by applying Gutzwiller projection to noninteracting mean-field states, but here we describe our unsuccessful attempts at doing so.  In our case, again referring to the two active bands as simply $s$ and $a$, we want to gap out the \emph{spin mode only} for only the symmetric band $s$.  A natural, potentially fruitful way to generalize the simple SBM is to add BCS pairing to the mean-field hopping Hamiltonian in Eq.~\eqref{eq:HMFhop}, $H_\mathrm{MF} \to H_\mathrm{MF} + \hat\Delta$, and project the mean-field ground state to $N_\mathrm{particles}=N_\uparrow+N_\downarrow=N_\mathrm{sites}$ total particles before Gutzwiller projection.  Working in momentum space, we could consider the following form for the pairing term:
\begin{equation}
\hat\Delta = \sum_k \left[\Delta_s f_{s,\uparrow}^\dagger(k)f_{s,\downarrow}^\dagger(-k) + \Delta_a f_{a,\uparrow}^\dagger(k) f_{a,\downarrow}^\dagger(-k) + \mathrm{H.c.} \right],
\end{equation}
where $f_{b,\alpha}^\dagger(k)$ creates single-particle states given by the wave functions in Eqs.~\eqref{eq:anti}--\eqref{eq:sym}.  Then by taking $\Delta_s\neq0$ and $\Delta_a=0$ we can selectively gap out band $s$ at the mean-field level.  However, doing so not only gaps out the corresponding spin mode (by pinning $\theta_{s\sigma}$), but it also disturbingly gaps out the corresponding charge mode (by pinning $\varphi_{s\rho}$).

To understand the latter, it is instructive to consider what happens when one adds BCS spin-singlet pairing to a single 1D band of spin-1/2 fermions and projects the ground-state wave function to $N$ total particles (at some generic density).  In this case, one will arrive at a BCS wave function with finite superconducting order parameter~(see, e.g., Ref.~\cite{Gros89_AnnPhys_189_53}), regardless of the fact that the Mermin-Wagner theorem prohibits such a ground state for a Hamiltonian that preserves particle number.  What is the fate of the system in terms of the bosonized fields?  The (singlet) superconducting pair operator reads
\begin{equation}
f_{R\uparrow}^\dagger f_{L\downarrow}^\dagger + f_{L\uparrow}^\dagger f_{R\downarrow}^\dagger \propto e^{-i\sqrt{2}\varphi_\rho}\cos(\sqrt{2}\theta_\sigma).
\end{equation}
This operator would take on a finite expectation value in the proposed wave function (in the sense of having finite two-point Cooper pair correlation functions at long distances).  Hence, both $\theta_\sigma$ and $\varphi_\rho$ would be pinned.  That is, we have constructed some pathological C0S0 state where the spin sector is indeed gapped, but the charge sector is ``soft'' ($g_\rho\to\infty$ in fact), as opposed to a bona fide C1S0 Luttinger liquid with finite $g_\rho$ (i.e., a Luther-Emery liquid).

\begin{figure}[t]
  \centering
  \includegraphics[width=0.7\columnwidth]{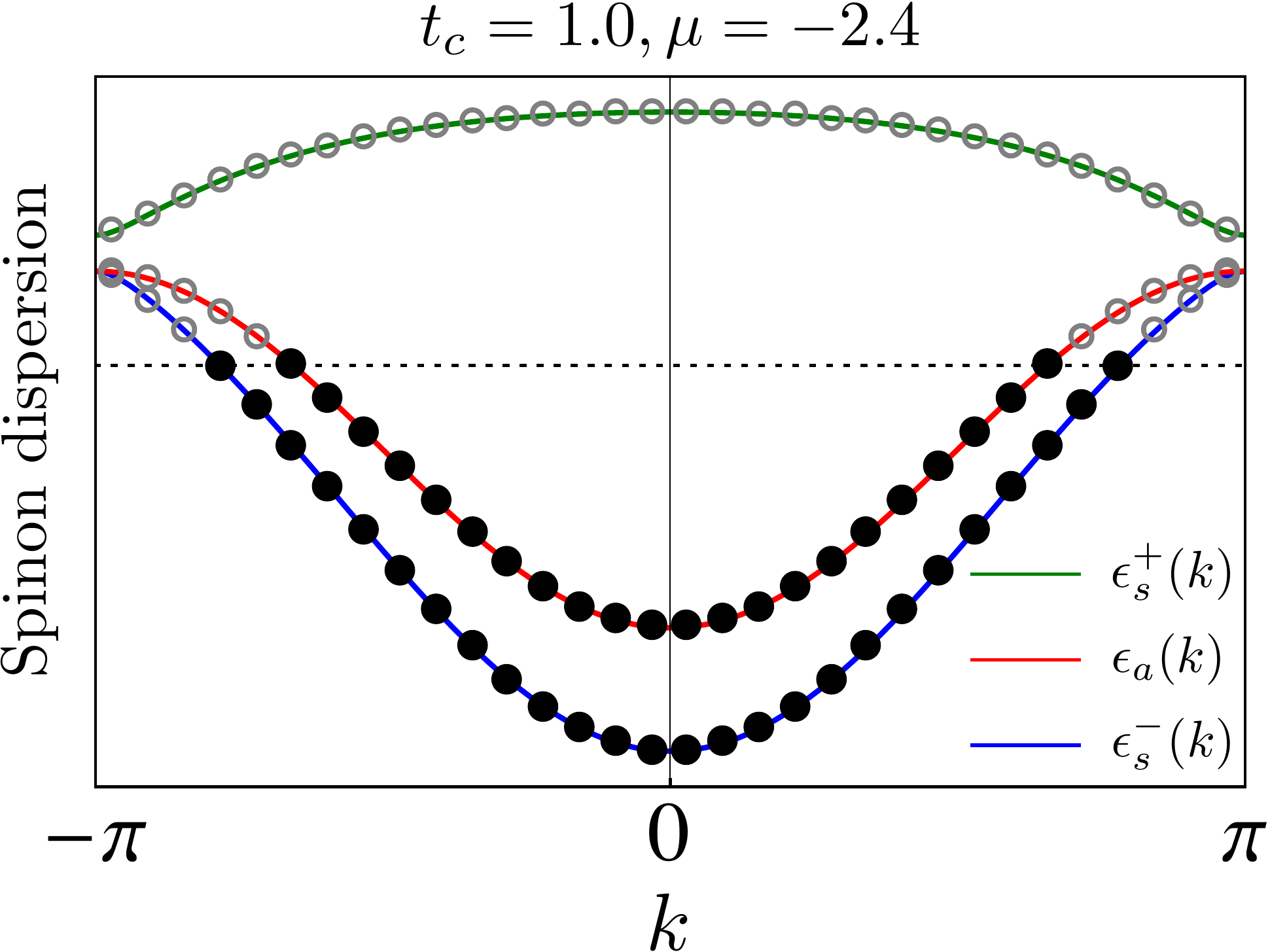}
  \caption{Specific SBM trial state used for the VMC data in Fig.~\ref{fig:ssf}.  The boundary conditions in the $x$ direction for the spinons are taken to be antiperiodic; this produces a spin wave function with periodic boundary conditions.}
  \label{fig:VMCstate}
\end{figure}

For the two-band situation on the kagome strip, at the mean field level upon taking $\Delta_s\neq0$ and $\Delta_a=0$,
we would therefore have pinned $\theta_{s\sigma}$ and $\varphi_{s\rho}$ fields.
Gutzwiller projecting the BCS wave function would then naturally simply pin the
remaining charge mode $\theta_{a\rho}$, thereby leaving a C0S1 state with
$c=1$.  The scaling dimensions of the bond-energy and spin operators with
respect to this fixed point are listed in the last row of Table~\ref{tab:2kFops} above. Insofar
as representing C1S1, this C0S1 BCS wave function is thus arguably
qualitatively worse than the C1S2 SBM wave function itself.  Most importantly,
the bond-energy at wave vector $q_<=2k_{Fs}$ is \emph{short-ranged} even at the
mean-field level (scaling dimension $\Delta=\infty$), whereas this is actually
the most prominent feature of the true C1S1 phase with its very slow power-law
decay ($\Delta=g_{\rho-}/4$). Given this catastrophic qualitative discrepancy,
we have not pursued numerical calculations of such BCS wave functions, and thus
must leave robust wave-function modeling of C1S1 for future work.

Returning to the SBM wave functions, we show in Fig.~\ref{fig:VMCstate} the
exact VMC state used for the spin structure factor calculations in
Fig.~\ref{fig:ssf} ($L=32$ PBC system with DMRG data taken at
at $J=0.9$).  Specifically, we choose $t_{c}=1.0, \mu=-2.4$, and antiperiodic
boundary conditions for the spinons in the $x$ direction.  This produces a
state whose $2k_{F}$ wave vectors match the singular features in the DMRG data.
Aside from having the extra feature in the spin structure factors at wave
vectors $q_<=2k_{Fs}$ (symmetric cases) and $\pi/2$ (antisymmetric case) as
well as exhibiting a quantitatively weak feature (in momentum space) in the
bond-energy at wave vector $q_<=2k_{Fs}$, such VMC states capture the
long-distance properties of the putative C1S1 phase reasonably well.  (The
relatively prominent feature shown by the VMC state at wave vector
$q_<=2k_{Fs}$ in the ``middle-middle'' structure factor
$\langle\mathbf{S}_{q}^{M}\cdot\mathbf{S}_{-q}^{M}\rangle$ is likely some
nonuniversal property of the given projected wave function; recall this feature
will be eliminated entirely in a true C1S1 state.)

\begin{figure}[t]
  \centering
  \includegraphics[width=0.85\columnwidth]{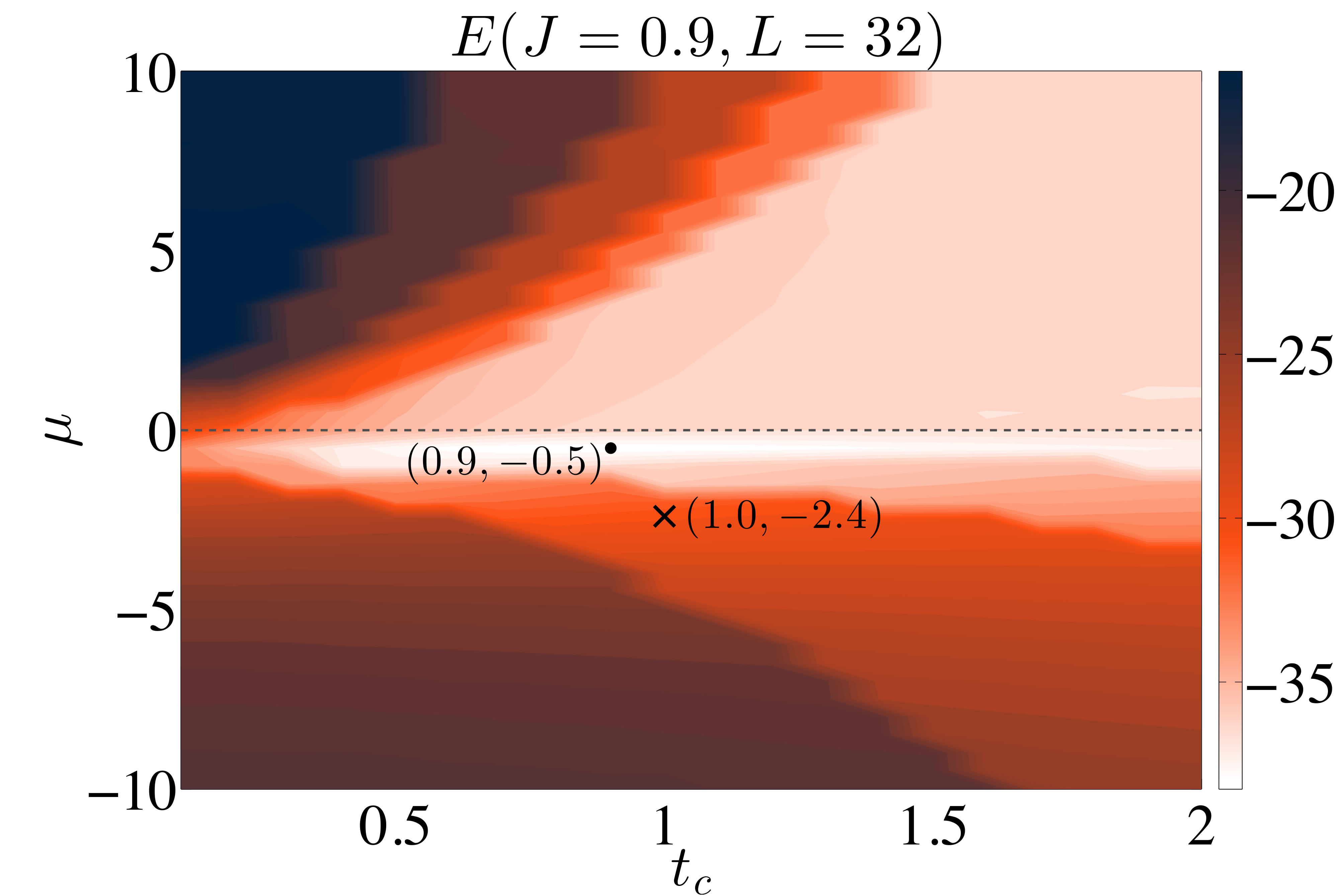}
  \caption{Energy landscape of SBM trial states versus $t_{c}$ and $\mu$ for a $L=32$ PBC system at $J=0.9$.  The point marked by $\,\boldsymbol{\cdot}\,$ is the energy-optimized state, while the point marked by $\,\boldsymbol{\times}\,$ is the state shown in Fig.~\ref{fig:ssf} (see also Fig.~\ref{fig:VMCstate}).}
  \label{fig:landscape}
\end{figure}

Finally, we discuss the energetics of our simple SBM trial states in the kagome
strip Heisenberg model; for concreteness, we continue to focus on the point
$J=0.9$ as in Fig.~\ref{fig:ssf}.  Within this class of SBM
states, the state at $t_{c}=1.0, \mu=-2.4$ shown in Fig.~\ref{fig:VMCstate} is
not quite the energy-optimized VMC state.  However, the lowest-energy
variational state is not far off at $t_{c}=0.9, \mu=-0.5$ [see
Fig.~\ref{fig:landscape} for the energy landscape at $J=0.9$ of our SBM trial
states in the variational space $(t_{c}, \mu)$].  This latter state has
incorrect values of $k_{Fs}$ and $k_{Fa}$ however (error $\sim 2\pi/L$). As for
the energies themselves, on the length $L=32$ PBC system at $J=0.9$, the DMRG
ground state has energy $-39.8$ (in units of the leg coupling $J_\ell$).  On the other hand, the energy-optimized VMC
state ($t_{c}=0.9, \mu=-0.5$) has energy $-38.3$, while the state chosen for
presentation ($t_{c}=1.0, \mu=-2.4$) has energy $-30.6$ (this can be improved somewhat by tuning $t_c$ and $\mu$ at fixed values of $k_{Fs}$ and $k_{Fa}$, e.g., $t_{c}=0.9, \mu=-1.5$ gives energy $-32.0$).  However, the latter is likely due to the state having inaccuracies in its (nonuniversal) amplitudes and short-range properties.  It should be possible to improve this deficiency by, for example, using the ``improved Gutzwiller'' wave functions of Ref.~\cite{Sheng09_PRB_79_205112}; these are essentially Gutzwiller-projected fully gapless superconducting wave functions, although empirically even they only have tunable amplitudes with fixed Luttinger parameter $g_{\rho-}=1$.  Even more importantly, recall that such SBM trial states are not even in the correct quantum phase (C1S2 instead of putative C1S1), so extremely accurate energetics should not be anticipated.

We emphasize again that the VMC wave functions are mainly meant to serve as a numerical representation/cross-check of the analytic parent C1S2 theory, as opposed to being quantitatively accurate trial states to describe all (including short-distance) properties of the DMRG data.  Still, our simple VMC states do reasonably well qualitatively, even semiquantitatively, with regards to those universal features shared between C1S2 and C1S1. \\ \\ \\ \\

\bibliography{kagome_strip}

\end{document}